\documentclass[journal]{IEEEtran}
\usepackage{makeidx}
\pdfoutput=1
\usepackage[backend=biber,bibstyle=ieee,citestyle=numeric-comp]{biblatex}

\usepackage{breakurl}
\usepackage{acronym}
\usepackage{multirow}
\usepackage{adjustbox}
\usepackage[shortlabels]{enumitem}
\usepackage{graphicx}
\usepackage{amsmath}
\usepackage{amssymb}
\usepackage{subcaption}
\usepackage{color}
\usepackage{soul}
\usepackage{listings,multicol}  
\usepackage{todonotes}
\usepackage{filecontents}
\usepackage{breqn}
\usepackage[font=sc]{caption}
\usepackage{colortbl}
\usepackage{enumitem}
\usepackage[ruled]{algorithm2e}
\usepackage{threeparttable}
\usepackage{lipsum}
\usepackage[mathlines,switch]{lineno}
\usepackage{booktabs}
\usepackage{textcomp}
\usepackage{pifont}
\usepackage{soul}
\usepackage{bm}
\acrodef{DP}    {data point}
\acrodef{PVS}   {Processed video sequence}
\acrodef{A/V}	{Audio/Video}
\acrodef{AOM}	{Alliance for Open Media}
\acrodef{AVC}	{Advanced Video Coding}
\acrodef{AWS}	{Amazon Web Services}
\acrodef{BD-BR}{Bjontegaard-Delta Bitrate}
\acrodef{bpc}	{bits per channel}
\acrodef{BQoE}	{Behavior Quality of Experience}
\acrodef{CBR}   {Constant Bitrate}
\acrodef{CABAC}	{Context-Adaptive Binary Arithmetic Coding}
\acrodef{CAVLC}	{Context-Adaptive Variable-Length Coding}
\acrodef{CDN}	{Content Delivery Network}
\acrodef{CDR}	{Continuous Dynamic Range}
\acrodef{CfE}	{Call for Evidence}
\acrodef{CIE}	{International Commission on Illumination}
\acrodef{CQoE}	{Cognitive Quality of Experience}
\acrodef{CRF}	{Constant Rate Factor}
\acrodef{CTU}	{Coding Transform Unit}
\acrodef{DASH}	{Dynamic Adaptive Streaming over HTTP}
\acrodef{DLM}	{Detail Loss Metric}
\acrodef{DR}    {Dynamic Range}
\acrodef{ECDF}  {Empirical Cumulative Distribution Function}
\acrodef{ECG}	{Electrocardiogram}
\acrodef{EEG}	{Electroencephalogram}
\acrodef{EPG}	{Electronic Program Guide}
\acrodef{EVC}   {MPEG-5 Essential Video Coding}
\acrodef{FHD}	{Full HD}
\acrodef{GOP}	{Group of Pictures}
\acrodef{GPU}	{Graphics Processing Unit}
\acrodef{HAS}	{HTTP Adaptive Streaming}
\acrodef{HD}	{High Definition}
\acrodef{HDR}	{High Dynamic Range}
\acrodef{HEVC}	{High Efficiency Video Coding}
\acrodef{HLS}	{HTTP Live Streaming}
\acrodef{HM}	{HEVC Test Model}
\acrodef{HLG}	{Hybrid Log-Gamma}
\acrodef{HLS}	{HTTP Live Streaming}
\acrodef{HVS}	{Human Visual System}
\acrodef{HW}	{Hammerstein-Wiener}
\acrodef{IP}	{Internet Protocol}
\acrodef{IPR}	{Intellectual Property Rights}
\acrodef{IQA}	{Image Quality Assessment}
\acrodef{JEM}   {Joint Exploration Test Model}
\acrodef{JND}	{Just Noticeable Difference}
\acrodef{JVET}  {Joint Video Exploration Team}
\acrodef{LF}    {Light Field}
\acrodef{LTE}	{Long Term Evolution}
\acrodef{MOS}	{Mean Opinion Score}
\acrodef{MPEG}	{Moving Pictures Expert Group}
\acrodef{MSE}   {Mean Square Error}
\acrodef{MV}	{Motion Value}
\acrodef{NARX}	{Nonlinear Autoregressive Network with Exogenous Inputs}
\acrodef{OTT}	{Over The Top}
\acrodef{OS}	{Operating System}
\acrodef{PLCC}	{Pearson Linear Correlation Coefficient}
\acrodef{PLR}	{Packet Loss Rate}
\acrodef{POP}	{Point of Presence}
\acrodef{PQ}	{Perceptual Quantizer}
\acrodef{PSNR} {Peak Signal to Noise Ratio}
\acrodef{QoE}	{Quality of Experience}
\acrodef{QoS}	{Quality of Service}
\acrodef{QP}	{Quantization Parameter}
\acrodef{RAN}	{Radio Access Network}
\acrodef{RD}	{Rate Distortion}
\acrodef{RMSE}	{Root Mean Square Error}
\acrodef{ROI}	{Region of Interest}
\acrodef{RR}	{Reduced Reference}
\acrodef{RTMP}	{Real Time Messaging Protocol}
\acrodef{RTT}	{Round Trip Time}
\acrodef{SAO}	{Sample Adaptive Offset}
\acrodef{SC}	{Spatial Complexity}
\acrodef{SD}	{Standard Definition}
\acrodef{SDR}	{Standard Dynamic Range}
\acrodef{SI}	{Spatial Information}
\acrodef{SoP}	{Sense of Presence}
\acrodef{SQI}	{Streaming Quality Index}
\acrodef{SROCC}	{Spearman's Rank Correlation Coefficient}
\acrodef{SSIM}	{Structural Similarity}
\acrodef{STSQ}	{Short Term Subjective Quality}
\acrodef{SVC}	{Scalable Video Coding}
\acrodef{SVM}	{Support Vector Machine}
\acrodef{SVR}	{Support Vector Regression}
\acrodef{TCP}	{Transmission Control Protocol}
\acrodef{TC}	{Temporal Complexity}
\acrodef{TI}	{Temporal Information}
\acrodef{TMO}	{Tone Mapping Operator}
\acrodef{TVSQ}	{Time Varying Subjective Quality}
\acrodef{UHD}	{Ultra High Definition}
\acrodef{UDP }	{User Datagram Protocol}
\acrodef{VCI} 	{Video Complexity Index}
\acrodef{VCEG}  {Video Coding Experts Group}
\acrodef{VMAF}  {Video Multimethod Assessment Fusion}
\acrodef{VoD}	{Video on demand}
\acrodef{VQA}	{Video Quality Assessment}
\acrodef{VQM}	{Video Quality Metrics}
\acrodef{VQMT}	{Video Quality Measurement Tool}
\acrodef{VBR}	{Variable Bit Rate}
\acrodef{VIF}	{Visual Information Fidelity}
\acrodef{VIFP}	{Visual Information Fidelity - Pixel Domain}
\acrodef{VR} 	{Virtual Reality}
\acrodef{VVC}   {Versatile Video Coding}
\acrodef{WCG}	{Wide Color Gamut}

\begin{document}

\title{Bjøntegaard Delta (BD): A Tutorial Overview of the Metric, Evolution, Challenges, and Recommendations}

\author{ Nabajeet~Barman,~\IEEEmembership{Member,~IEEE,} Maria G.~Martini,~\IEEEmembership{Senior~Member,~IEEE}, Yuriy Reznik,~\IEEEmembership{Senior~Member,~IEEE} 
\thanks{Copyright © 20xx IEEE. Personal use of this material is permitted. However, permission to use this material for any other purposes must be obtained from the IEEE by sending an email to pubs-permissions@ieee.org.}
\thanks{Nabajeet Barman is with Sony Interactive Entertainment, London, UK. Yuriy Reznik is with Brightcove, Boston, US. Maria Martini is with Wireless Multimedia \& Networking Research Group, Kingston University London, UK. Part of this work was done during Nabajeet's previous employment at Brightcove, London, UK. \newline
Author emails: n.barman@ieee.org, m.martini@kingston.ac.uk, and yreznik@brightcove.com}}



\maketitle

\begin{abstract}
The Bjøntegaard Delta (BD) method proposed in 2001 has become a popular tool for comparing video codec compression efficiency. It was initially proposed to compute bitrate and quality differences between two Rate-Distortion curves using PSNR as distortion metric. Over the years, many works have calculated and reported BD results using other objective quality metrics such as SSIM, VMAF and, in some cases, even subjective ratings (mean opinion scores). However, the lack of consolidated literature explaining the metric, its evolution over the years, and a systematic evaluation of the same under different test conditions can result in a wrong interpretation of the BD results thus obtained.

Towards this end, this paper presents a detailed tutorial describing the BD method and example cases where the metric might fail. We also provide a detailed history of its evolution, including a discussion of various proposed improvements and variations over the last 20 years. In addition, we evaluate the various BD methods and their open-source implementations, considering different objective quality metrics and subjective ratings taking into account different RD characteristics. Based on our results, we present a set of recommendations on using existing BD metrics and various insights for possible exploration towards developing more effective tools for codec compression efficiency evaluation and comparison. 
\end{abstract}

\begin{IEEEkeywords}
Video Compression, Codec Comparison, Bjøntegaard Delta, BD-Rate, BD-PSNR, Rate–distortion 
\end{IEEEkeywords}

\section{Introduction}

Recent years have seen a rise in media consumption enabled by over-the-top video streaming services such as Netflix, YouTube, and Twitch, with global IP video traffic currently comprising approximately 82\% of all IP traffic by 2022, up from 75\% in 2017~\cite{cisco_forecast}. The increasing popularity of such services can be attributed primarily to increased network bandwidth, better compression efficiency, and the proliferation of 
video playback devices such as smartphones, tablets, and internet-connected TVs. To cater to the increased user expectation of any content anytime, anywhere, and on any device, there has been a recent rise of many alternative technologies being developed and offered in the market. Given the plethora of new technologies, it can often be confusing for a service provider to select the best for their usage. 

Among critical choices a streaming provider has to make is adopting a video coding technology to efficiently reduce a media file size for faster transmission and delivery over the network. However, with an increasing number of new codecs being developed, it is often difficult for the industry to decide which one to choose. In addition to the cost, many other factors must be considered for adopting a new codec. Such factors include consideration of compression efficiency gain, speed, cost, application, existing support on devices, and number of encodes required/tolerated. Of all these factors, one of the most critical factors to be considered for adopting a new codec is the compression efficiency gains provided over its predecessor and competitors. 

Over the past 20 years, an increasing number of new codecs have been developed to achieve higher compression efficiency without loss of perceived quality. These range from H.264/AVC~\cite{h264} published as a standard in 2003 to HEVC~\cite{h265}, published in 2013, to more recently published codec such as VVC~\cite{h266}, LCEVC~\cite{lcevc}, and AV1~\cite{av1}. During the development of a standard, many new coding tools are proposed. The proposed coding tools are evaluated for their performance, usually in terms of compression efficiency gain and complexity. If the proposed tools are found to have a positive impact (e.g., in terms of gain considering compression vs. complexity trade-off) on the overall coding framework, they are included in the draft standard. The development of such newer video coding standards remains one of the primary forces behind the increasing rise and popularity of streaming services.
 
One of the most commonly used methods for calculating the difference in compression efficiency between two codecs or encoding modes is using bitrate and quality savings as calculated using the Bjøntegaard Delta (BD) method first proposed by Gisle Bjøntegaard in 2001~\cite{VCEG-M33}. The BD method was initially proposed to compare two versions of the same video encoder with different video coding tools switched on/off~\cite{HSTP-VID-WPOM}. An example of its usage in the standardization process in the past includes a comparison of the different versions of reference software, such as between reference software for HEVC, HM (HEVC Test Model) 6.0 vs. HM 7.0 after the addition of new coding tools~\cite{JCTVC-L1100}. Another common usage includes comparing two different generations of codecs, for example, performance evaluation of the Test Model 5 of Versatile Video Coding (VTM 5.0) over HM 16.20 in different configuration modes~\cite{JVET-O0003}. 

The BD metric was initially designed and used to compare \ac{RD} curves for fixed Quantization Parameter (QP) based encoded video sequences. In the fixed QP-based rate control, the QP parameters are set constant for the whole video sequence, and the same level of compression is applied to all macroblocks of a frame. Such QP-based encoding usually ensures “good” overlap between the two compared RD curves, which are said to be “well-behaved.” For example, as discussed in~\cite{HSTP-VID-WPOM}, the common practice is to compress each test sequence using four different QP values (usually 22, 27, 32, and 37). After encoding, 
PSNR 
and bitrate 
are calculated for each encoded sequence, and then the BD-Rate and BD-PSNR values are calculated. Such 4-point testing provides sufficient information about each codec's performance, allowing their RD curves to have good overlaps\footnote{By overlap here we refer to the common quality and bitrate range resulting in a good common integration area between the compared two RD curves as shown later in Figure~\ref{fig:bd-psnr}.} 
in their ranges of bitrates and PSNR levels. In some cases, the RD curves may overlap or cross over, resulting in unique instances in which the results thus obtained might be unreliable.

\subsection{Motivation}

Over the years, the BD metric has been used to compute coding efficiency gains using RD curves obtained for settings and metrics other than what was originally used in the metric design. For example, as discussed earlier, the usual practice in standardization activities is to use BD measurements for fixed QP-based encoded sequences. Using constant QP-based rate control and QP-based sampling ensures that the bitrates and PSNR values at each point are monotonically increasing. This allows simple interpolation and integration techniques to be used, as essential for computation of BD-Rate or BD-PSNR values~\cite{VCEG-M33}. Since then, many works~\cite{GuoNetflix2018PCS,Katsenou2019CodecComparison,Li2019CodecComparison,DarkhanAV1} have compared RD curves for videos encoded using other rate control methods such as constant quality or constrained bitrate encoding~\cite{CHEN200719}. Such cases might result in not-so-well-behaved RD curves with a good overlap in the values range (see Section~\ref{sec:overlapRange} and Section~\ref{sec:overlapMetric})
which can often result in unpredictable results.

Additionally, one can observe the use of quality metrics other than PSNR for BD computation. PSNR, the metric used in the design of BD calculations, remains the choice of quality metric to measure distortion in standardization activities. However, many other (non-standardization) works have used other objective quality metrics such as \ac{SSIM} \cite{SSIM} and \ac{VMAF} \cite{NetflixVMAF_Github}, and in some cases, subjective measurements such as \ac{MOS} \cite{barman2022lcevc,LorenzoMHV22, ugur2010high} as metrics to measure the amount of distortion introduced. Nevertheless, there exist
quite a lot of differences in the behavior of the metrics. For example, PSNR is unbounded while other metrics such as SSIM and VMAF are bounded. Also, there is a high saturation of SSIM at higher bitrates. Additionally, one of the issues with using
subjective quality ratings in terms of  \ac{MOS} is that they are not always monotonically increasing, which can result in unreliable BD results~\cite{VCEG-AL23, scenic, barman2022lcevc,pezzulli2020estimation}. Therefore, the suitability of alternative metrics for BD measurements is not well established.  

\subsection{Objectives of this Work}

A systematic evaluation of the BD metric and its variations, given the era of newer codecs and quality metrics, is still missing. Hence,  the reliability of such measurements and results, mainly when not supported with RD plots or raw measurement values, remains an open question. Also, many works have indicated that the magnitude of savings obtained using the BD metric is often quite different compared with the results obtained from subjective experiments~\cite{barman2022lcevc,scenic,Gary2016BDBR}. This raises many open questions that we try to address in this work. 

A better understanding of the BD metric, its evolution and variants, and special cases where it might result in unreliable values can help design alternative measurement metrics for other content and applications such as immersive videos, video coding for machines and \ac{HDR}. In the special case of adaptive bitrate streaming and cases when there is information about network and understanding of probabilities of different operating points, and hence their relative importance for overall comparison, one can compute much more relevant assessment of codec performance compared to the uniform-density average as currently done. The main objective of this paper is to help the general audience understand the BD metric theory, its evolution, and special cases where the results should be interpreted with caution. Additionally, based on results obtained from an experimental study, we provide a set of recommendations on the best practices for the computation of coding compression efficiency.

\subsection{Related Work}

Since its initial proposal, many variations of the BD metric have been proposed over the years. In the same period, many MS Excel, Python, and MATLAB-based open-source implementations of the BD metrics have been made available. Both the industry and academia widely use such open-source implementations. However, in the absence of documentation, it is not always clear which BD method the respective implementation provides. However, for some cases, depending on the nature of RD curves, the values reported by the different variants of the BD method could be rather different~\cite{bd-br-MHV}. Hence such cross-comparison of results across different works can be misleading and not lead to a fair comparison of the video codecs. 

To address the lack of appropriate literature providing a theoretical understanding of the BD method, an ITU-T technical paper was published in July 2020, which describes BD computation for video coding experiments~\cite{HSTP-VID-WPOM}. However, the scope of the work was limited only to a conceptual level overview of the metrics, reasons behind some of the choices, references to technical papers, and a discussion of some situations where the results should be interpreted cautiously. However, no evaluation was performed, and no “recommendations'' as such were provided. 

\begin{table}[t!]
\caption{Commonly used acronyms.}
\def\arraystretch{1.2}
\centering
\begin{tabular}{|ll|}
\hline
\multicolumn{2}{|c|}{\textbf{Compression Performance Evaluation}}                       \\ \hline
\multicolumn{1}{|l|}{BD-BR}      & Bjøntegaard Delta BitRate                      \\ \hline
\multicolumn{1}{|l|}{BD-Rate}    & Bjøntegaard Delta Rate                         \\ \hline
\multicolumn{1}{|l|}{BD-PSNR}    & Bjøntegaard Delta Quality                      \\ \hline
\multicolumn{1}{|l|}{DR (or D/R)}         & Distortion Rate                                \\ \hline
\multicolumn{1}{|l|}{RD (or R/D)}         & Rate Distortion                                \\ \hline
\multicolumn{2}{|c|}{\textbf{Quality Metrics}}                                          \\ \hline
\multicolumn{1}{|l|}{MOS}              & Mean Opinion Score                             \\ \hline
\multicolumn{1}{|l|}{PSNR}             & Peak Signal-to-Noise Ratio                     \\ \hline
\multicolumn{1}{|l|}{SSIM}             & Structural Similarity                          \\ \hline
\multicolumn{1}{|l|}{VMAF}             & Video Multimethod Assessment Fusion            \\ \hline
\multicolumn{2}{|c|}{\textbf{Video Compression Standards}}                              \\ \hline
\multicolumn{1}{|l|}{H.264/AVC}        & Advanced Video Coding                          \\ \hline
\multicolumn{1}{|l|}{H.265/HEVC}       & High Efficiency Video Coding                   \\ \hline
\multicolumn{1}{|l|}{H.266/VVC}        & Versatile Video Coding                         \\ \hline
\multicolumn{2}{|c|}{\textbf{Standards Organization}}                                   \\ \hline
\multicolumn{1}{|l|}{ISO}              & International Organization for Standardization \\ \hline
\multicolumn{1}{|l|}{IEC}              & International Electrotechnical Commission      \\ \hline
\multicolumn{1}{|l|}{ITU}              & International Telecommunication Union          \\ \hline
\multicolumn{1}{|l|}{JVET}             & Joint Video Experts Team                       \\ \hline
\multicolumn{1}{|l|}{JVT}              & Joint Video Team                               \\ \hline
\multicolumn{1}{|l|}{JCT-VC}           & Joint Collaborative Team on Video Coding       \\ \hline
\multicolumn{1}{|l|}{VCEG}             & Video Coding Experts Group                     \\ \hline
\end{tabular}
\label{tab:abbreviations}%
\end{table}

\subsection{Contributions}
Considering the above-described reasoning and objectives, we present in this paper the following contributions:
\begin{enumerate}
    \item A tutorial on the BD metric, its formulation, detailed history of its evolution and its usage in the standardization efforts and academic publications. 
    \item A discussion of special cases where the results obtained using BD measurements might be unreliable.
    \item A detailed evaluation of the different variations of BD functions and existing open-source implementations for their correctness using an open-source dataset. 
    \item An evaluation of the performance of the BD metric (and its variations) across different objective quality metrics.
    \item A performance evaluation of the BD metric using subjective scores as such RD curves are not always well-behaved. A comparison with an alternative metric is also presented.
    \item A set of recommendations on the best practices and open-source implementations to use.
    \item A discussion on the possible extensions of the BD metric.
\end{enumerate}


\subsection{Outline}
The rest of the paper is organized as follows. Section~\ref{sec:bdbr_Theory} presents the tutorial explaining the calculation of BD metrics, BD-Rate and BD-PSNR. Section~\ref{sec:crossoverillus} presents a detailed discussion of special cases where the BD metric might produce not so accurate results. Section~\ref{sec :bdbr_History} presents the history of the evolution of the BD metric and its different variants, along with a brief review of its use in various academic and industrial publications. An experimental study of different implementations and variants of the BD metric is presented in Section~\ref{sec:ExpStudy} with Section~\ref{sec:obser} presenting the discussions, key observations, and recommendations based on the results from the experimental study and literature review. Section~\ref{sec:possibleExtension} presents a discussion on the possible future work towards increased applicability of the BD metric given newer quality metrics and applications. Section~\ref{sec:conclusion} concludes the paper. For easier understanding and reference, Table~\ref{tab:abbreviations} presents a list of commonly used abbreviations in this work.

\begin{figure}[t!]
\begin{center}
\includegraphics[width=0.9\linewidth,height=0.4\textwidth]{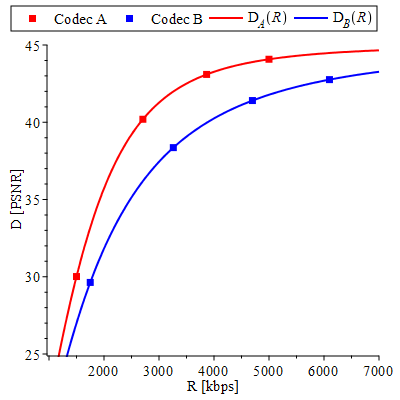}
\caption{Distortion-Rate performance comparison of two codecs, codec A and codec B. Red points show the measured (R,D) operating points for codec A. Blue points show the measured (R,D) operating points for codec B. The functions $D_A(R)$ and $D_B(R)$ show the results of interpolation across sample points and extrapolation beyond. These functions can be understood as approximations of Operational Rate-Distortion characteristics of codecs A and B, respectively.}
\label{fig:bdbr-conceptual}
\end{center}
\end{figure}

\section{ Main principles of codec performance comparison and BD metrics} \label{sec:bdbr_Theory}

\subsection{RD Performance of Codecs}

Rate-distortion (RD) theory~\cite{Shannon-RD,Shannon-RD2, Berger1971RateDT, davissonRD} provides foundation for developing lossy data compression algorithms. RD theory is primarily concerned with finding the most compact representation of a stochastic source, subject to a fidelity criterion~\cite{Ortega-RD}. The RD functions provided by this theory define the fundamental trade-off between encoding rate and distortion characteristics that may be reachable in practice.


The distortion measurement in the image and video compression field 
has been primarily done using the objective quality metric \ac{PSNR}, primarily due to its simplicity and ease of computation and relation to MSE (mean square error) metric, used to arrive at many important results in classic rate distortion theory.


\subsection{RD Curves}

For a given codec, if we consider the possible quantization choices, we can define an operational rate-distortion curve which is obtained by plotting the distortion achieved by the particular codec for each rate. The choice of quantization parameters (QP) values is based on the most commonly anticipated operational range of the proposed codec's applications. Over the years, RD curves plotted for two or more codecs became the de-facto representation for studying the trade-off between different coding tools and codecs. Consider the Distortion-Rate performance for two codecs, Codec A and Codec B. Let, ${A_1, A_2, ..., A_N}$ and ${B_1, B_2, ..., B_N}$ be the measured $N$ operating points for Codec A and Codec B respectively. Then, one can define the measured distortion values, $D[PSNR]$, for Codec A and Codec B as $[D_{A_1}, < D_{A_2}, < ..., < D_{A_N}]$ and $[D_{B_1}, < D_{B_2}, < ..., < D_{B_N}]$ respectively. The corresponding bitrate values for Codec A and Codec B can be defined as  $[R_{A_1}, < R_{A_2}, < ..., < R_{A_N}]$ and $[R_{B_1}, < R_{B_2}, < ..., < R_{B_N}]$ respectively. 

Figure~\ref{fig:bdbr-conceptual} presents conceptual Distortion-Rate curves for performance comparison of two codecs, codec A and codec B, with distortion measured using PSNR considering $N$= 4 (R,D) operating points. Looking at Figure~\ref{fig:bdbr-conceptual}, one can easily conclude that codec A is better than codec B, as it delivers higher PSNR at the same bitrate value(s). However, to say how much better it is, one needs to specify the bitrate for the comparison, e.g., $R_x$, and then look at the difference
\begin{equation}
  \Delta_D(R_x) = D_A(R_x) – D_B(R_x). 
\end{equation}
This is very simple to compute, but the result is not a single value, as $\Delta_D(R_x)$ becomes a function of the reference rate point $R_x$. We can also repeat the same with respect to attempts to quantify the differences in rates:
\begin{equation}
    \Delta_R(D_x) = R_A(D_x) – R_B(D_x),
\end{equation}
where $R_A(D)$ and $R_B(D)$ are the inverse functions for $D_A(R)$ and $D_B(R)$ respectively. However, $\Delta_R(D_x)$ is also not a single value, but rather a function of the reference distortion point, $D_x$. This inconvenience of direct differences subsequently led to the necessity of quantifying the “average" performance differences between two RD curves leading to the creation of performance measurement methods known today as Bjøntegaard Delta (BD) methods.

\subsection{Average Performance Metrics: BD-Rate and BD-PSNR}

In April 2001, Gisle Bjøntegaard submitted a contribution “VCEG-M33”~\cite{VCEG-M33} in the ITU-T SG16 Q.6 13\textsuperscript{th} VCEG meeting in Austin, Texas, USA, where a method to provide relative gain between two methods by measuring the average difference between the two RD curves was proposed. The basic proposal was to fit a third-order (cubic) polynomial curve through 4 data points and then find an expression for the integral of the curve. The average bitrate “savings”, referred to as BD-Rate, was then calculated as the difference between the integrals divided by the integration interval. Since the higher bitrates in a “normal” RD plot dominated the bitrate savings, it was proposed to take the logarithm of the bitrates, resulting in $dB$ units on both axes. Using the logarithm of bitrate values also allowed for the “reciprocity” of calculation of change in bitrate or change in PSNR, thus allowing for calculation of both Quality (PSNR) savings and Bitrate savings. The quality savings is referred to as BD-PSNR. The fundamental elements for estimating the Bjøntegaard Deltas (BD-Rate and BD-PSNR) as proposed in~\cite{VCEG-M33} can be summarized as follows:
\begin{enumerate}
    \item Fit a curve through 4 data points.
    \item Based on the curve fitting, find an expression for the integral of the curve.
    \item The average difference is then calculated as the difference between the integrals divided by the integration interval.
\end{enumerate}

In the original contribution, PSNR-bitrate values are assumed to be obtained for QP values 16, 20, 24, and 28. However, in more recent codec development (e.g., HEVC and VVC), with new test conditions (increased resolution and high frame rate videos), different QP values (e.g., 22, 27, 32, and 37) have since then been used to obtain respective PSNR-bitrate values.

\subsection{BD-PSNR Calculation} \label{sec:bd-psnr}

\begin{figure}[t!]
\begin{center}
\includegraphics[width=0.9\linewidth,height=0.4\textwidth]{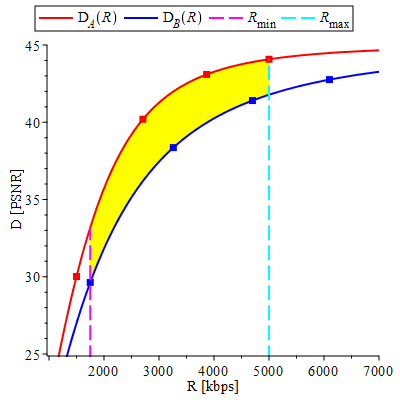}
\caption{ Computation of BD-PSNR. $R_{min}$ and $R_{max}$ indicate the range of integration along bitrate, and the yellow region shows that the area of the integral: $A_D= \int_R [D_A(R)-D_B(R)] dR$. The average BD-PSNR value in dB is computed as BD-PSNR, $\bar{\Delta}_D$ = $A_D/(R_{max}-R_{min})$ [dB].}
\label{fig:bd-psnr}
\end{center}
\end{figure}

Figure~\ref{fig:bd-psnr} presents an illustration of the calculation of average BD-PSNR ($\bar{\Delta}_D$) considering RD curves for two codecs, codec A and codec B. The Y-axis denotes distortion ($D$) measured using PSNR, while the X-axis represents the bitrate R (in kbps). The common overlap between two RD curves is shown using the dashed lines, with $R_{max}$ and $R_{min}$ denoting the upper and lower bound, respectively, where $R_{min} = max(R_{A_1}, R_{B_1})$ and $R_{max} = min(R_{A_N}, R_{B_N})$. After finding the overlapping region between the two RD curves, the expression for the integral for the two RD curves is obtained using a curve fit. The integration 
\begin{equation}
   \int_{R_{min}}^{R_{max}} \Delta(R) dR 
\end{equation}
 is then performed over the common overlapping area (using lower and upper bound, $R_{min}$ and $R_{max}$ respectively). As mentioned earlier, in the VCEG-M33 contribution, a third-order polynomial (cubic) fit for curve fitting was used. However, since then, other fitting functions have been proposed~\cite{COM-16C404,beyondBDBR}. The shaded area in yellow captures the integral difference between the two $D(R)$ curves:
\begin{equation}
    A_D= \int_R [D_A(R)-D_B(R)] dR.
\end{equation}
The average BD-PSNR (dB) value computed over the common integration area, $\bar{\Delta}_D$ is then computed as 
\begin{equation}
    \frac{A_D}{(R_{max}-R_{min})}.   
\end{equation}

\subsection{BD-Rate Calculation}

\begin{figure}[t!]
\begin{center}
\includegraphics[width=0.9\linewidth,height=0.4\textwidth]{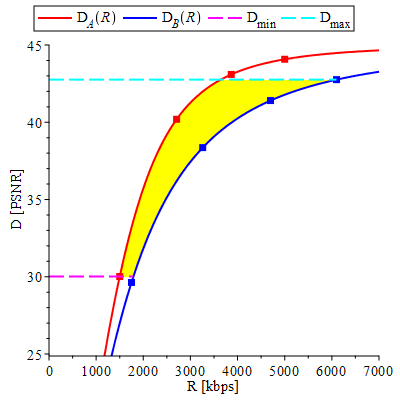}
\caption{Computation of BD-Rate. $D_{min}$ and $D_{max}$ indicate the range of integration along distortion (PSNR), and the yellow region shows 
the area of the integral: $A_R= \int_D [R_A(D)-R_B(D)] dD$. The average BD-Rate value is then computed as  $\bar{\Delta}_R =  \frac{A_R}{(D_{max}-D_{min})},$}
\label{fig:bd-rate}
\end{center}
\end{figure}

Similar to Figure~\ref{fig:bd-psnr}, Figure~\ref{fig:bd-rate} presents an illustration of the calculation of BD-Rate ($\bar{\Delta}_R$) considering RD curves for two codecs, codec A and codec B. The Y-axis denotes distortion, $D$ measured using PSNR, while the X-axis represents the bitrate, R (in kbps). The common overlap between the two RD curves in terms of distortion is shown using the dashed lines, with $D_{max}$ and $D_{min}$ denoting the upper and lower bound, respectively, where, $D_{min} = max(D_{A_1}, D_{B_1})$ and $D_{max} = min(D_{A_N}, D_{B_N})$. After finding the overlapping region between the two RD curves, the expression for the integral for the two RD curves is obtained using a curve fit. The integration 
\begin{equation}
    \int_{D_{min}}^{D_{max}} \Delta(D) dD
\end{equation}
is then performed over the common overlapping area (using lower and upper bound, $D_{min}$ and $D_{max}$ respectively). The shaded area in yellow indicates the savings figure, BD-Rate calculated as
\begin{equation}
    \bar{\Delta}_R = \frac{A_R}{(D_{max}-D_{min})},  
\end{equation} 
where
\begin{equation}
  A_R = \int_D [R_A(D)-R_B(D)] dD.  
\end{equation}

\subsection{Discussion}

The Bjøntegaard Delta method, as proposed in \cite{VCEG-M33} uses a logarithmic scale for the domain of the bitrate interpolation. Hence, considering that $R$ values are in $log_{10}$ scale, the percentage bitrate savings BD-Rate (\%) can be expressed as, 
\begin{equation}
   100 \cdot (10^{\bar{\Delta}_R} - 1). 
\end{equation}
Due to ease of interpretation of the percentage bitrate savings for equal measured quality, BD-Rate (\%) is more commonly used and reported than BD-PSNR (dB). Using BD-Rate savings figures also makes it easier to compare savings figures for two RD curves across different quality metrics.

\begin{figure}[t!]
\begin{center}
\includegraphics[width=0.9\linewidth,height=0.4\textwidth]{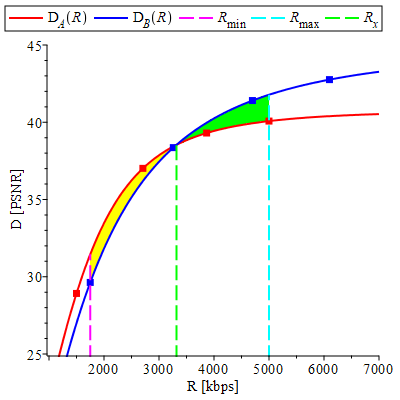}
\caption{Example sample case with crossover RD curves when using BD metric may be confusing.}
\label{fig:bdbr-crossover}
\end{center}
\end{figure}
\begin{figure*}[t!]
    \centering
    \begin{subfigure}[t]{0.31\textwidth}
    \includegraphics[ width=1.0\textwidth]{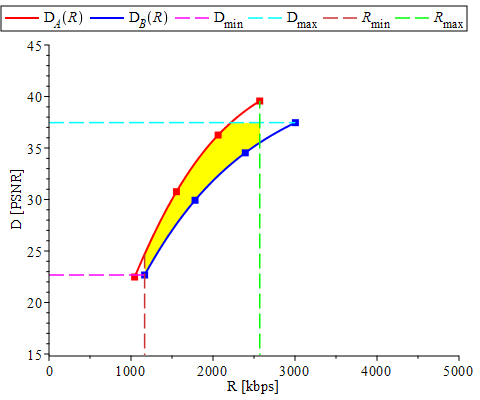}
    \caption{}
    \label{fig:5a}
    \end{subfigure}%
    \begin{subfigure}[t]{0.32\textwidth}
    \centering
    \includegraphics[ width=1.0\textwidth]{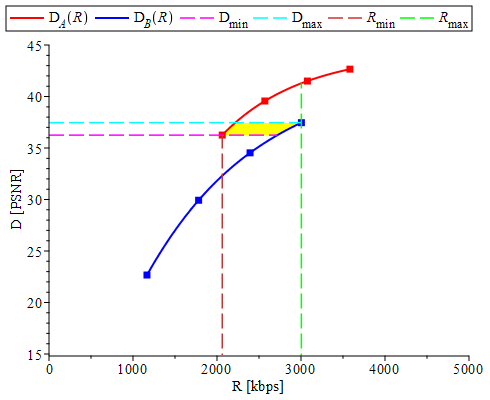}
    \caption{}
    \label{fig:5b}
    \end{subfigure}
    \begin{subfigure}[t]{0.32\textwidth}
    \includegraphics[ width=1.0\textwidth]{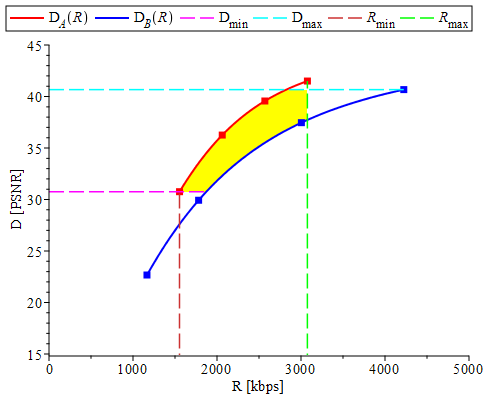}
    \caption{}
    \label{fig:5c}
    \end{subfigure}%
    \caption{Illustration showing example cases with different overlap cases between the RD curves. (a) Well-behaved case with a good overlap between the RD curves. (b) Too ``low" case with a small overlap area between the RD curves. (c) Too ``high" case with too large of overlap between the RD curves than what the Quality-Bitrate values are obtained for Codec A.}
\label{fig:bdbr-BRoverlap}
\end{figure*}
\section{Special Cases} \label{sec:crossoverillus}

While average bitrate ($\bar{\Delta}_R$) and PSNR savings ($\bar{\Delta}_D$), discussed previously, may simplify expressions of relative gains, in some cases reliance only on such metrics can lead to wrong conclusions. In this section, we will present and highlight some cases where BD metrics might fail and should be used with caution. Although such special cases where the values obtained using the BD metric might be unreliable are well known among the experts involved in standardization activities, it is not so well known among other users, due to the lack of user guides and relevant literature. This can lead to possible misinterpretation of the reported BD-Rate and BD-Quality results\footnote{By BD-Quality, we refer to the average quality savings using any quality metric such as SSIM and VMAF (not limited to PSNR)}.

\subsection{Cross Over Between the RD Curves}

One prominent example case where the BD metric can fail is when there is a crossover between the RD curves. Such cases are commonly observed in many real-world applications, especially when using \ac{MOS} scores as the choice of measurement of signal distortion \cite{bonnineauBDCorssOver,barman2022lcevc}. Such an example case is illustrated in Figure~\ref{fig:bdbr-crossover} where there is a crossover between the RD curves, $D_A(R)$ and $D_B(R)$ for two codecs, codec A and codec B, respectively. In this case, the D/R curves for both codecs $D_A(R)$ and $D_B(R)$ intersect at some intermediate point 
\begin{equation}
      R=R_x, \quad R_{min} \leq R_x \leq R_{max}.
\end{equation}
Consequently, the difference between $D_A(R)$ and $D_B(R)$ in the range $[R_{min},R_x)$ will have the opposite sign compared to the differences in the range $(R_x,R_{max}]$. Hence, the value of full integral in $[R_{min},R_{max}]$ will be the difference between the areas shown and yellow and green in Figure~\ref{fig:bdbr-crossover}. These areas could be close, thus resulting BR value close to 0, while, at the same time, it clear that in the range $R<R_x$ codec A is better, while in the range $R>R_x$ codec B becomes superior. This shows that BD metrics must always be used with caution, and in case of intersecting behaviors, one should look at the application and identify a range of prime interest. 

\begin{figure}[t!]
\begin{center}
\includegraphics[width=0.9\linewidth,height=0.4\textwidth]{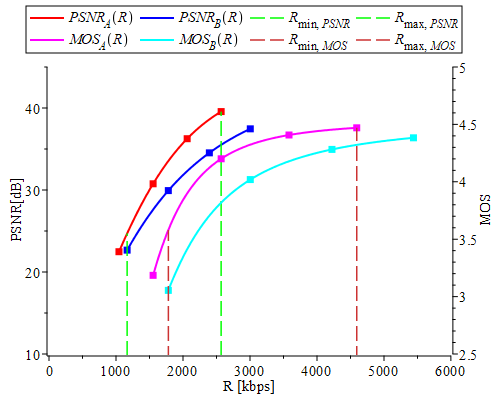}
\caption{Illustration showing an example case where depending on the considered quality metric, different overlapping bitrate ranges (BitrateRange-MOS and BitrateRange-PSNR) are considered for BD metric calculations.}
\label{fig:bdbr-rangemetric}
\end{center}
\end{figure}
\subsection{Overlapping Bitrate Range and/or Quality Range} ~\label{sec:overlapRange}
\begin{table*}[t!]
  \centering
  \caption{List of Major Standardization Contributions Related to BD Metric.}
    \def\arraystretch{1.12}
    \resizebox{1.0\linewidth}{!}{
    \begin{tabular}{|l|p{6.4em}|p{18.1em}|p{22.5em}|c|}
    \hline
    \multicolumn{1}{|p{2.9em}|}{\textbf{Year}} & \textbf{Contribution} & \textbf{Title} & \textbf{Description} & \multicolumn{1}{p{4.8em}|}{\textbf{Reference}} \\
    \hline
    2001  & VCEG-M33 & Calculation of average PSNR differences between RD-curves & First contribution proposing the method to calculate Rate and PSNR savings  & \cite{VCEG-M33} \\
    \hline
    2007  & VCEG-AE07 & An excel add-in for computing Bjontegaard metric and its evolution & Provides an Excel Add-in computing the BD-BR metric as defined in VCEG-M33 & \cite{VCEG-AE07} \\
    \hline
    2008  & VCEG-AI11 & Improvements of the BD-PSNR model & Improvements of the earlier BD-PSNR model to compute gain at low rates or high rates  & \cite{VCEG-AI11} \\
    \hline
    2008  & ITU COM16-C.404 & On the calculation of PSNR and bit-rate differences for the SVT test data & First mention about unexpected results for the BD-Rate metric due to cubic fitting when evaluating the ultra high definition sequences and proposal for using piecewise cubic polynomial interpolation. & \cite{COM-16C404} \\
    \hline
    2009  & VCEG-AL22 & Reliability metric for BD measurements & Makes some additional suggestions to make sure that BD measurements are accurate & \cite{VCEG-AL22} \\
    \hline
    2009  & VCEG-AL23 & BD measurements based on MOS & MOS based results (since they are not always monotonous) can be unreliable and hence BD-BR measurements should be limited to only PSNR & \cite{VCEG-AL23} \\
    \hline
    2009  & ITU COM16–C358–E & Improvements of Excel macro for BD-gain computation & Proposes an improvement of the previously defined macro in VCEG-AE07 and reliability metrics defined in VCEG-AL22 & \cite{COM-16C358E} \\
    \hline
    2011  & - & Excel template for BD-rate calculation based on Piece-wise Cubic Interpolation & Excel add-on implementing Piecewise Cubic Hermite Polynomial Interpolation & \cite{bossen_piecewise} \\
    \hline
    2017  & JVET-H0030 & BD-Rate/BD-PSNR Excel extensions & Some extensions to the JCT-VC/JVET Excel template for the computation of BD-rate and BD-PSNR numbers along with support for more than 4 data points. & \cite{JVET-H0030} \\
    \hline
    2019  & JCTVC-J0003 & JVET AHG report: Test model software development (AHG3) & BD-rate calculation based on Piecewise Cubic Interpolation for VVC standardization work. & \cite{JVET-O0003} \\
    \hline
    \end{tabular}}
  \label{table:contributions}
\end{table*}
Fixed QP encoding that is traditionally used in standardization activities results in monotonically increasing quality-bitrate encodes and usually ensures a good overlapping (integration) area between the compared RD curves. However, if the measured codecs are over two generations apart or when considering encoding settings other than fixed QP (such as CRF or VBR encoding), as is common in many non-standardization activities, quality-bitrate points for the two RD curves can be quite different. This can lead to different overlapping areas where the BD metric computation occurs. Figure~\ref{fig:bdbr-BRoverlap} illustrates three example cases considering PSNR as the quality metric. It can be observed that in Fig. 5(b) and Fig. 5(c), the BD-Rate and BD-Quality results do not necessarily represent the average coding efficiency for all the measured RD points for Codec A and Codec B as compared to what is observed in the ``well-behaved" case shown in Fig. 5(a). Considering that most works only report BD-Rate (or BD-Quality) results and not detailed RD curves denoting the overlapping ranges, such savings figures might lead to an inaccurate interpretation of the performance of the compared codecs. Hence, for more realistic savings figures, there should be a good overlap between the compared RD curves, with the overlap between the RD curves covering the quality-bitrate ranges suitable for the considered application \cite{Gary2016BDBR}. 


\subsection{Different Overlapping (Integration) Bitrate Ranges Among Metrics} ~\label{sec:overlapMetric}

The overlapping bitrate ranges might differ depending on the quality metric used for BD-Rate computation. Such cases are more common when encoding settings other than fixed QP (e.g., constant quality) are used to obtain quality-bitrate points. Such an example case is illustrated in Figure~\ref{fig:bdbr-rangemetric} where we can observe that the bitrate range considered for BD measurements, \textit{BitrateRange-MOS} and \textit{BitrateRange-PSNR}, can be pretty different across the two quality metrics, PSNR and MOS in this example case. As discussed by the authors in \cite{Gary2016BDBR}, since the results obtained using one metric can be applied to another different scenario, it is vital to understand the different ranges on which they have been computed. Thus, a comparison of BD-Quality and BD-Rate values across different metrics should be made with caution. Alternatively, a possible practical solution to generalize and compare results obtained using different metrics would be to consider a bitrate range where the compared RD curves have a good common overlap for the compared quality metrics. BD values obtained thus will result in more realistic savings figures.

\section{History of Development of Bjøntegaard Delta and its Evolution} \label{sec :bdbr_History}

The origin, evolution, and usage of BD metric can be traced to standardization activities for the development of various video compression standards such as H.264/AVC, H.265/HEVC, and more recently, H.266/VVC ~\cite{h264,h265,h266}. Hence, a brief discussion of various standardization groups and activities is provided next to help better understand the development and use of the BD metric in various standardization activities, followed by a detailed discussion on the evolution of the BD metric over the past 20 years. We end the section with a review of literature using the BD metric. 

\subsection{Standardization Groups}

ITU-T Video Coding Experts Group (VCEG) is the informal name for the International Telecommunication Union - Telecommunication (ITU-T) Study Group 16 Question 6\footnote{https://www.itu.int/en/ITU-T/studygroups/2017-2020/16/Pages/q6.aspx} where work on visual coding within the ITU is undertaken. ISO/IEC JTC1 SC29\footnote{https://committee.iso.org/home/jtc1sc29} is the Subcommittee of the Joint Technical Committee of the International Organization for Standardization (ISO) and the International Electrotechnical Commission (IEC), which focuses on coding of audio, picture, multimedia and hypermedia information. Since 2001, ITU-T Study Group 16 (VCEG) and ISO/IEC JTC1 SC29/ WG11 (MPEG) have jointly created different groups to develop various video coding standards. Joint Video Team (JVT) was created in 2001, resulting in ITU-T Rec. H.264/AVC~\cite{h264} followed by the Joint Collaborative Team on Video Coding (JCT-VC) in 2010, resulting in the development of a new coding standardization, ITU-T H.265/HEVC~\cite{h265}. More recently, Joint Video Experts Team (JVET) was created on 27 October 2017, resulting in the development of a new video compression standard, ITU-T H.266/VVC.

\begin{table*}[t!]
\def\arraystretch{1.3}
  \centering
  \caption{Summary of some recent works on video codec comparison which used BD metrics. \small{Codecs compared, type of encoder used, evaluation methodology, quality metrics used to compare the efficiency and the focus applications, along with few observations, are presented in the table.}}  
  \resizebox{1.0\linewidth}{!}{
\begin{tabular}{|c|c|c|c|c|c|}
\hline
\textbf{Work}   & \textbf{Year} & \textbf{Codecs Compared} & \textbf{Quality Metrics Used} & \textbf{Video Resolutions Considered} & \textbf{Focus Application} \\ \hline
\cite{Barman2017QoMEX17} &
  2017 &
  H.264, H.265, VP9 &
  PSNR, SSIM, VIFp &
  upto 1080p &
  Live \\ \hline
  \cite{Topiwala2017CodecComparison360} & 2017          & AV1, HEVC, JVET          & SPSNR-NN, WS-PSNR, MOS        & 8/10-bit 8K 360   (Spherical) Video   & \textit{on-Demand}         \\ \hline
\cite{Topiwala2019CodecComparisonEVC} &
  2019 &
  VVC, AV1 and EVC &
  PSNR, MOS &
  8/10-bit, upto 4K &
  \textit{on-Demand} \\ \hline
\cite{Laude2018CodecComparison} &
  2018 &
  JEM, AV1, HEVC &
  PSNR &
  upto 4K &
  on-Demand \\ \hline
\cite{Pinar2018CodecComparison} &
  2018 &
  HEVC, VP9, AV1 &
  PSNR, MOS &
  720p &
  Broadcast \\ \hline
\cite{GuoNetflix2018PCS} &
  2018 &
  H.264, H.265, VP9,   AV1 &
  PSNR, VMAF &
  upto 1080p &
  on-Demand \\ \hline
\cite{Topiwala2018CodecComparison} &
  2018 &
  VVC, AV1 and HEVC &
  PSNR, MOS &
  upto 4K, both 8 and   10-bit &
  \textit{on-Demand} \\ \hline
\cite{Grois2018CodecComparison} &
  2018 &
  AV1, JEM, VP9, HEVC &
  PSNR &
  upto UHD, 360 &
  \textit{on-Demand} \\ \hline
\cite{Li2019CodecComparison} &
  2019 &
  AVC, HEVC, VP9, AVS2,   AV1 &
  MOS &
  upto UHD &
  \textit{on-Demand} \\ \hline
\cite{Katsenou2019CodecComparison} &
  2019 &
  AV1, HEVC &
  PSNR, VMAF, MOS &
  upto UHD, 10-bit &
  on-Demand \\ \hline

\cite{DarkhanAV1} &
  2020 &
  H.264, H.265 and AV1 &
  PSNR, SSIM, VMAF, MOS &
  upto FHD &
  Live \\ \hline
\cite{GamingHDRVideoSET} &
  2021 &
  H.264, H.265, VP9 and   AV1 &
  PSNR, HDR-VQM &
  4K &
  Live \\ \hline
\cite{grois_codec2021} &
  2021 &
  HEVC, EVC, VVC, AV1 &
  PSNR &
  4K &
  VoD \\ \hline
\cite{codec_4k8kHDR} &
  2021 &
  HEVC,    VVC, AV1 &
  \multicolumn{1}{l|}{PSNR, SSIM, and HDR-VDP} &
  4K and 8K &
  \textit{VoD} \\ \hline
\cite{barman2022lcevc} &
  2022 &
  H.264, H.265, LCEVC &
  PSNR and VMAF &
  FHD &
  Live \\ \hline
\end{tabular}
}
   \begin{tablenotes}
      \small
      \item EVC:  MPEG-5 Essential Video Coding, VVC: Versatile Video Coding, JVET: Joint Video Exploration Team, JEM: Joint Exploration Test Model, LCEVC: Low Complexity Enhancement Video Coding.
        \newline
        The text in italics font is deduced based on the encoding setting used in the paper and is not explicitly mentioned in the paper. 
        \newline 
        Works in \cite{Topiwala2018CodecComparison}, \cite{Topiwala2017CodecComparison360} and \cite{Topiwala2019CodecComparisonEVC} used informal MOS scores rather than MOS scores obtained from ITU-T recommended subjective test procedure.
    \end{tablenotes}

\label{tab:CodecComparisonLiterature}
\end{table*}
\subsection{Evolution of the BD Metric}

Soon after the initial proposal in~\cite{VCEG-M33}, BD\footnote{While so far we have used Bjøntegaard Delta (BD) to refer to the BD-Rate and BD-PSNR methods proposed in VCEG-M33~\cite{VCEG-M33}, it should be noted that in the contribution VCEG-M33, no formal name was used for the proposed methods for calculation of rate and quality savings. However, over the years, the term “Bjøntegaard Delta” became popular in the standards community. In most cases, it often only refers to commonly used BD-Rate savings.} metrics gained acceptance in the standardization community. As it is easier to interpret bitrate savings rather than quality savings, BD-Rate soon found its use in the standardization community. It has since then been used to select newer coding tools and compare the codec compression efficiency of a newly developed video compression standard with its predecessor (and competitors). Initially a software called \textit{avsnr4} was developed and made available in the h26L directory on the FTP site\footnote{ftp://ftp3.itu.int/video-site/H26L/avsnr4.zip}. The software reported two types of differences~\cite{VCEG-N81,VCEG-AA10}:
\begin{enumerate}
    \item Average difference in bitrate between two curves (BD-Rate) - measured in \%
    \item Average difference in PSNR between two curves (BD-PSNR) - measured in dB.
\end{enumerate}

\begin{figure*}[t!]
    \centering
    \begin{subfigure}[t]{0.45\textwidth}
    \includegraphics[ width=1.0\textwidth]{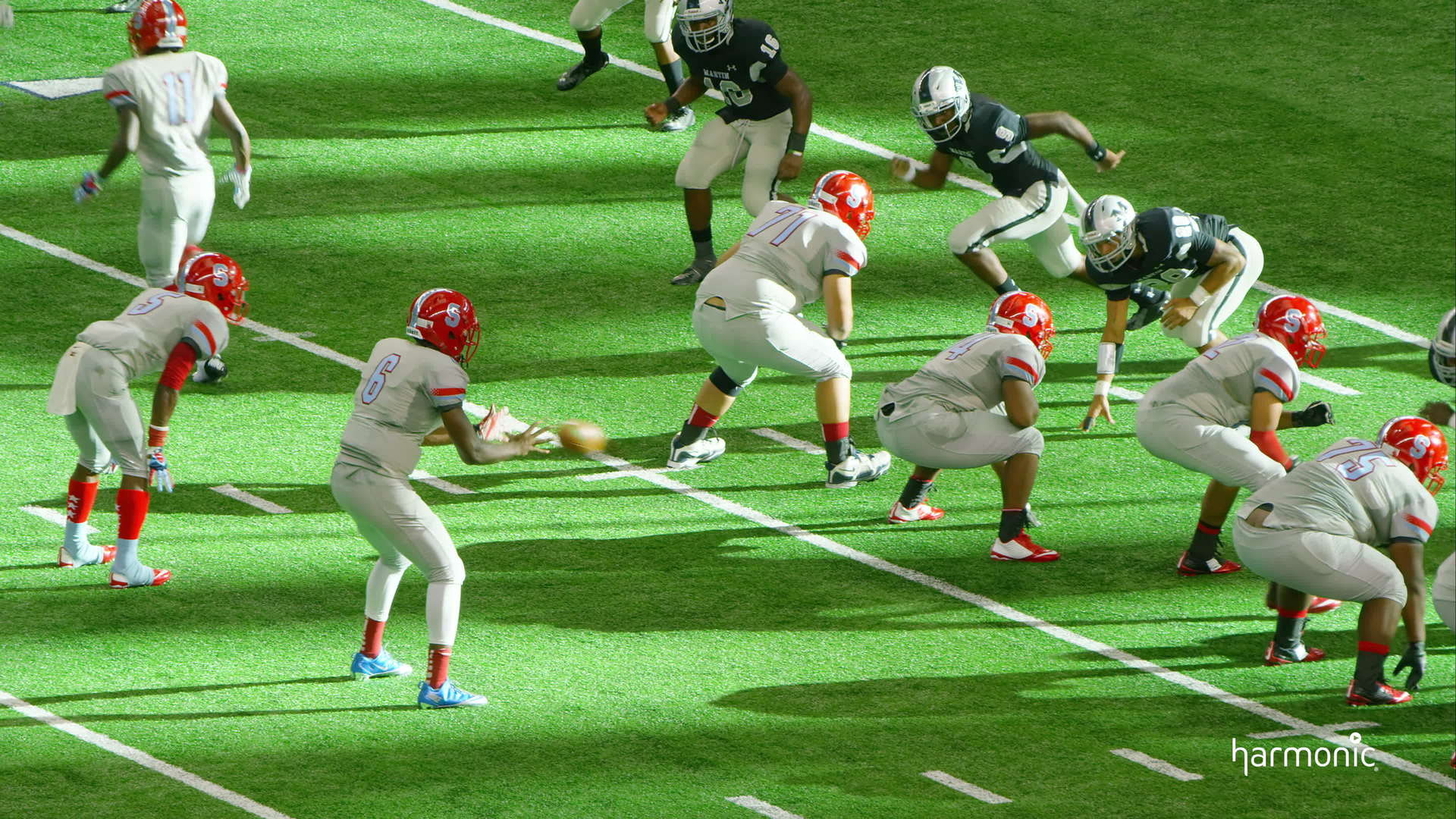}
    \caption{Video 1: American Football.}
    \label{fig:bar_x264}
    \end{subfigure}%
    \hspace{1cm}
    \begin{subfigure}[t]{0.45\textwidth}
    \centering
    \includegraphics[ width=1.0\textwidth]{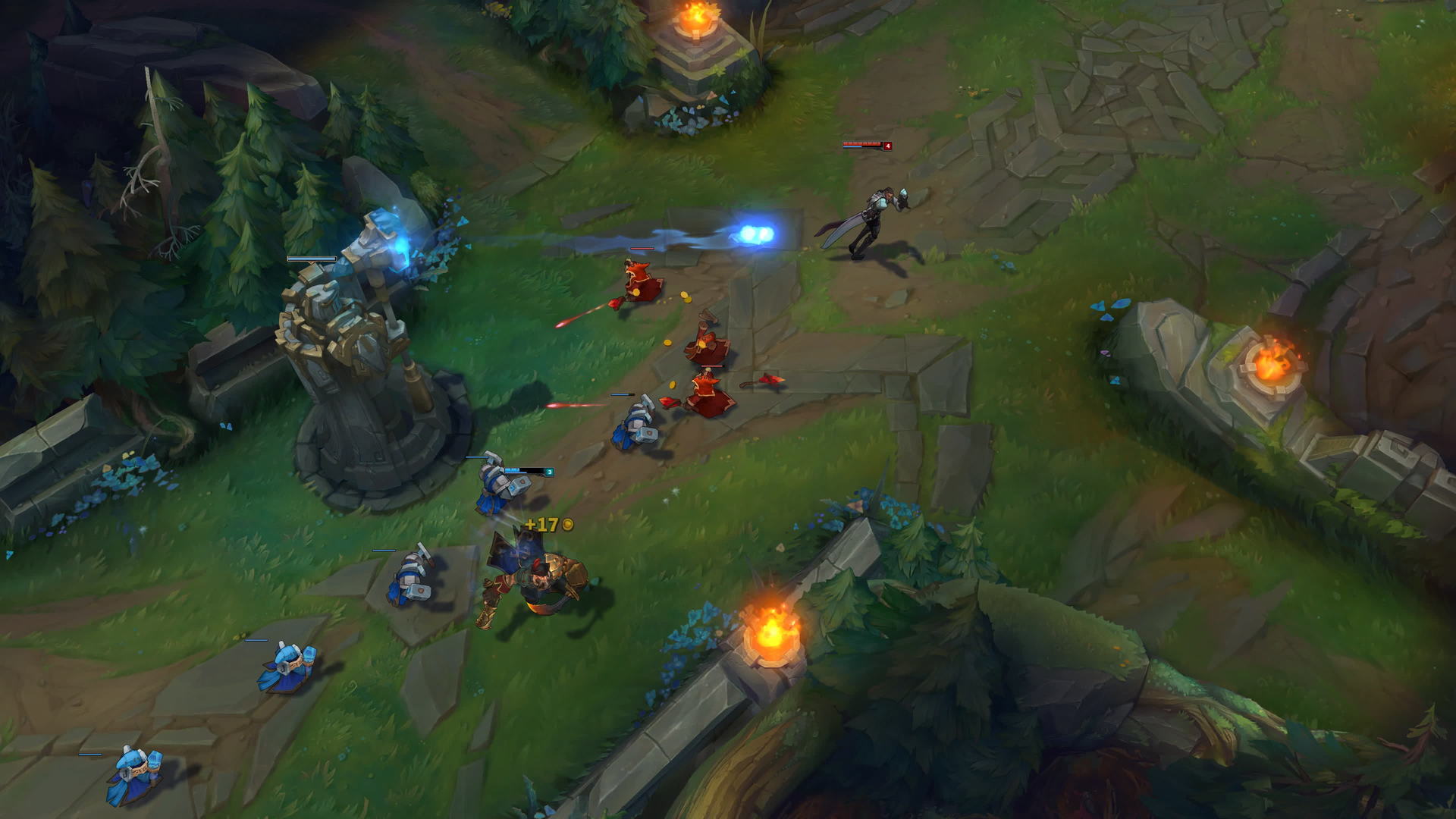}
    \caption{Video 2: League of Legends.}
    \label{fig:bar_x265}
    \end{subfigure}
    
    \begin{subfigure}[t]{0.45\textwidth}
    \includegraphics[ width=1.0\textwidth]{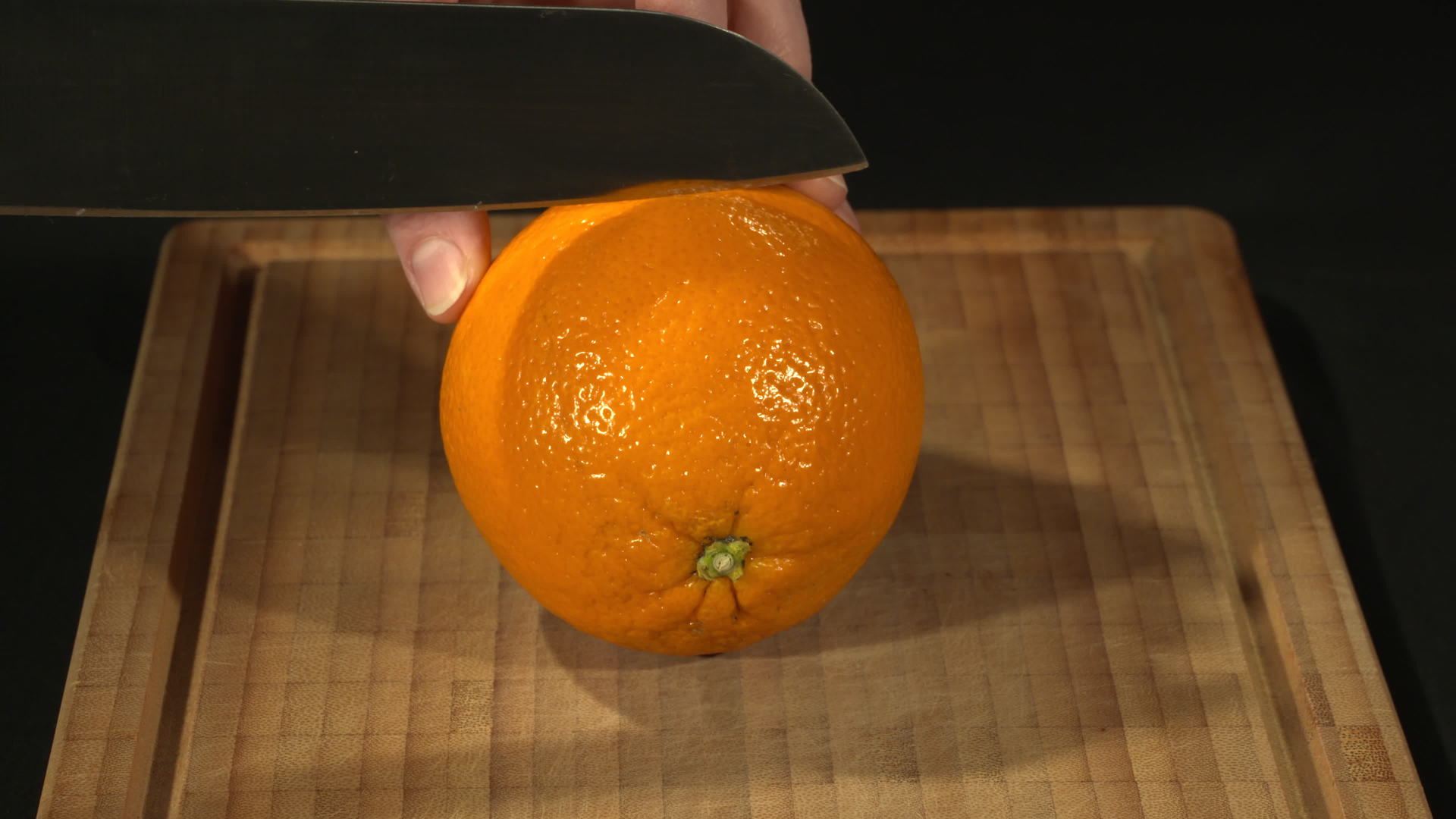}
    \caption{Video 3: Cutting Orange.}
    \label{fig:bar_x264}
    \end{subfigure}%
    \hspace{1cm}
    \begin{subfigure}[t]{0.45\textwidth}
    \centering
    \includegraphics[ width=1.0\textwidth]{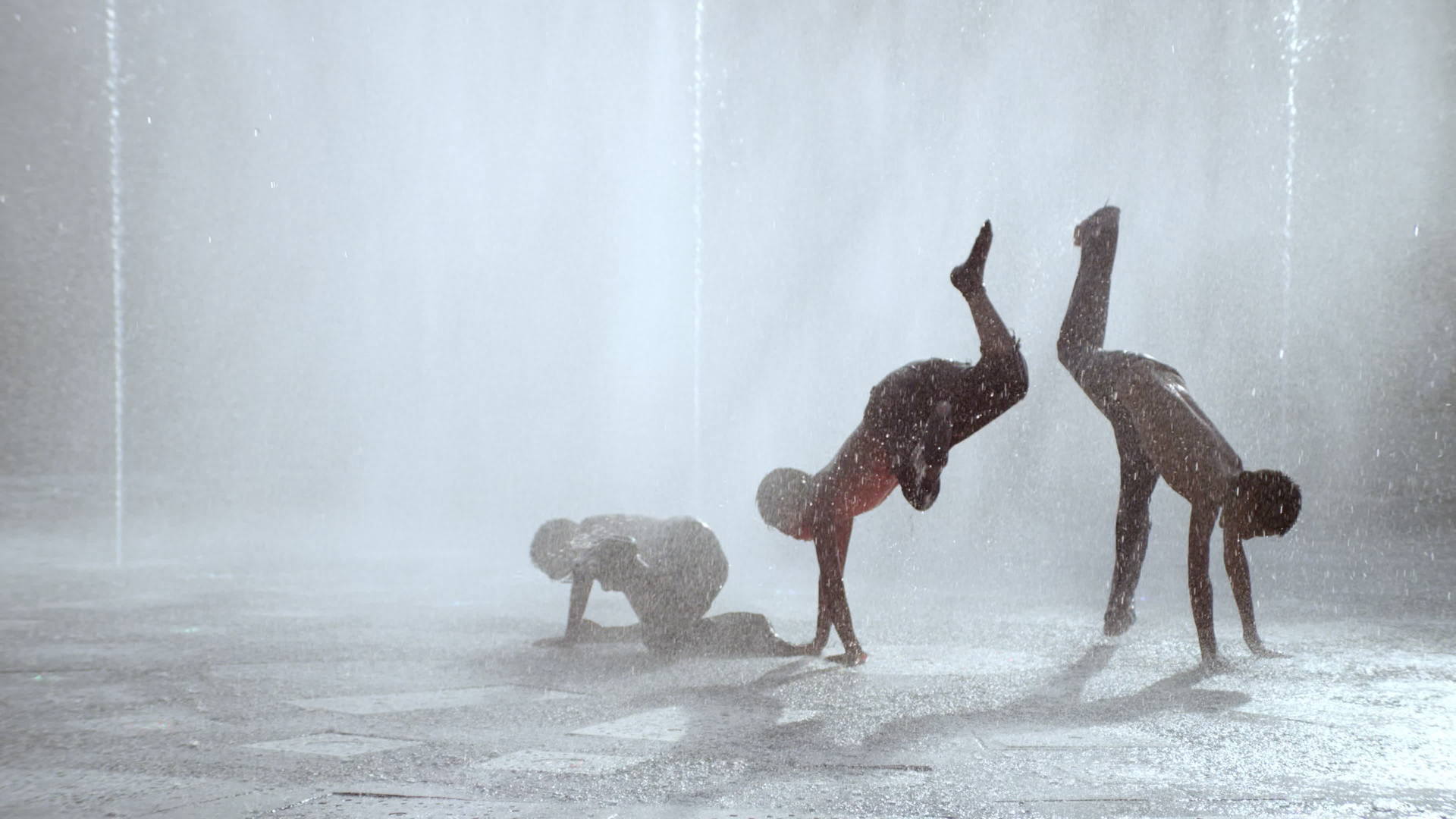}
    \caption{Video 4: Water.}
    \label{fig:bar_x265}
    \end{subfigure}
    \caption{Screenshots of the four video sequences considered in this work.}
    \label{fig:screenshots}
\end{figure*}
The proposed method and the sample implementation were since then used in standardization activities from the selection of various coding tools~\cite{JVT-B077} for evaluating the performance of the newly developed codecs using their reference implementations to their predecessors (e.g., H.26L vs. H.263~\cite{VCEG-AA10}). Over the years, different implementations and variations of the BD method started to appear. Table~\ref{table:contributions} presents the significant contributions related to the BD method in various standardization meetings summarizing the evolution of the BD metric. Next, we discuss some of the contributions.

In 2007, an Excel add-in that computes the BD-Rate and BD-PSNR metric as defined in VCEG-M33 was presented in the 31\textsuperscript{st} VCEG meeting~\cite{VCEG-AE07} followed by a more detailed contribution later in 2009~\cite{COM-16C358E}. In July 2008, Gisle provided a contribution to the improvement of BD-Rate, proposing an enhanced version of the Excel macro where, instead of average BD gains, it was proposed that for RD curves consisting of 5 or more data points, “BD-Rate low” and “BD-Rate high” 
could be computed, where low and high referred to low QP and high QP data points, respectively~\cite{VCEG-AI11}.

In April 2008, the first instance of the instability of cubic fitting was pointed out by Sharp in ITU contribution
COM16-C.404~\cite{COM-16C404}. The authors reported unexpected results for the BD-Rate metric when evaluating the ultra high definition sequences due to the use of cubic interpolation. Instead of cubic fitting, the authors proposed using the ``Piecewise Cubic Hermite Interpolating Polynomial" (henceforth piecewise-cubic), which was found to provide a better fit for the given data points. In July 2011, J. Wang et al. also identified issues with BD-Rate calculation for Class-A sequences as part of Common Test Conditions, used in the development of HEVC~\cite{JCTVC-F270}. The authors noted that the BD-Rate calculation based on piecewise cubic interpolation using the \textit{hm32piecewisecubic2.xls} Excel add-in appears to be a practical fix to the issue~\cite{bossen_piecewise}. During the JCTVC meeting, it was recommended to investigate further the use of the current BD-Rate tool in the evaluation of proposals. It was also suggested to report results for both cubic and piecewise-cubic interpolation~\cite{JCTVC-F800}. Since then, the standardization efforts for developing newer codecs have primarily been using the piecewise cubic fitting implementation (e.g., for developing HEVC standard and reference software implementation)~\cite{JCTVC-G1200, JVET-O0003, JCTVC-G003, JCTVC-L1100}.

Despite improvements over the years of the original method and its implementation(s), certain limitations existed. For example, the original implementations could only support four rate-distortion points. Also, a good overlap between the two curves (which is said to be a case of “well-behaved” RD curves) is a requirement to compute meaningful results. To address these gaps, in Oct 2017, Tourapis et al. in the 8\textsuperscript{th} JVET meeting, proposed a new contribution, JVET-H0030. In this contribution, they proposed a new Excel template for the calculation of BD values for more than 4 data points. They proposed various modes to the original function, which allows for calculating values by extrapolation (or interpolation) when the RD curves are not initially overlapping. The contribution includes an Excel template allowing for the computation of the old BD formulation and the proposed formulation. The ability to compute “region of interest” measurements is also provided. 
While 
during the discussion it was agreed to plan to use this in the call for proposals spreadsheet template~\cite{JVET-H_Notes_dA}, JVET still uses the \textit{hm32piecewisecubic2.xls} since they only deal with 4 points and their data is “well behaved”~\cite{JVET-O0003, JVET-N1010}. 


\subsection{On the Use of BD-Rate and BD-PSNR Metric in Literature}

Over the years, many works other than in the standardization community have used BD measurements to evaluate the performance of different codecs and coding tools. Table~\ref{tab:CodecComparisonLiterature} presents few such works on codec compression efficiency comparison. It can be observed that, while initially proposed and commonly used with PSNR as the objective quality metric to measure the distortion, over the years BD-Rate and BD-PSNR methods have since been used with other quality metrics (e.g., SSIM, VMAF, and MOS). In addition, other metrics have also been used when considering other video content, such as \ac{HDR} and immersive videos. However, no systematic investigation has been done for the evaluation of the performance of the BD metric given newer quality metrics or other contents. As discussed in ~\cite{HSTP-VID-WPOM}, the proposed metric has been designed for 2D, 8-bit content, and its use for other quality metrics and content should be done with caution. Therefore, in the next section, through different studies, we evaluate the performance of the BD metrics and their variations and different open-source implementations on an open-source dataset representing different RD characteristics. For clarity and readability, when using quality metrics other than PSNR for BD measurements, we will use the following notations:

\hspace{\parindent} BD-Rate (Quality Metric), e.g., BD-Rate (SSIM) indicating percentage bitrate savings when using SSIM \cite{SSIM} for distortion measurement.

\hspace{\parindent} BD-Quality (Quality Metric): e.g., BD-Quality (SSIM) indicating quality savings in terms of SSIM scores.

Unless mentioned otherwise, in the rest of this paper, the BD-Rate value refers to percentage bitrate savings using PSNR as a distortion metric.

\begin{figure*}[t!]
\begin{center}
\includegraphics[width=1.0\linewidth]{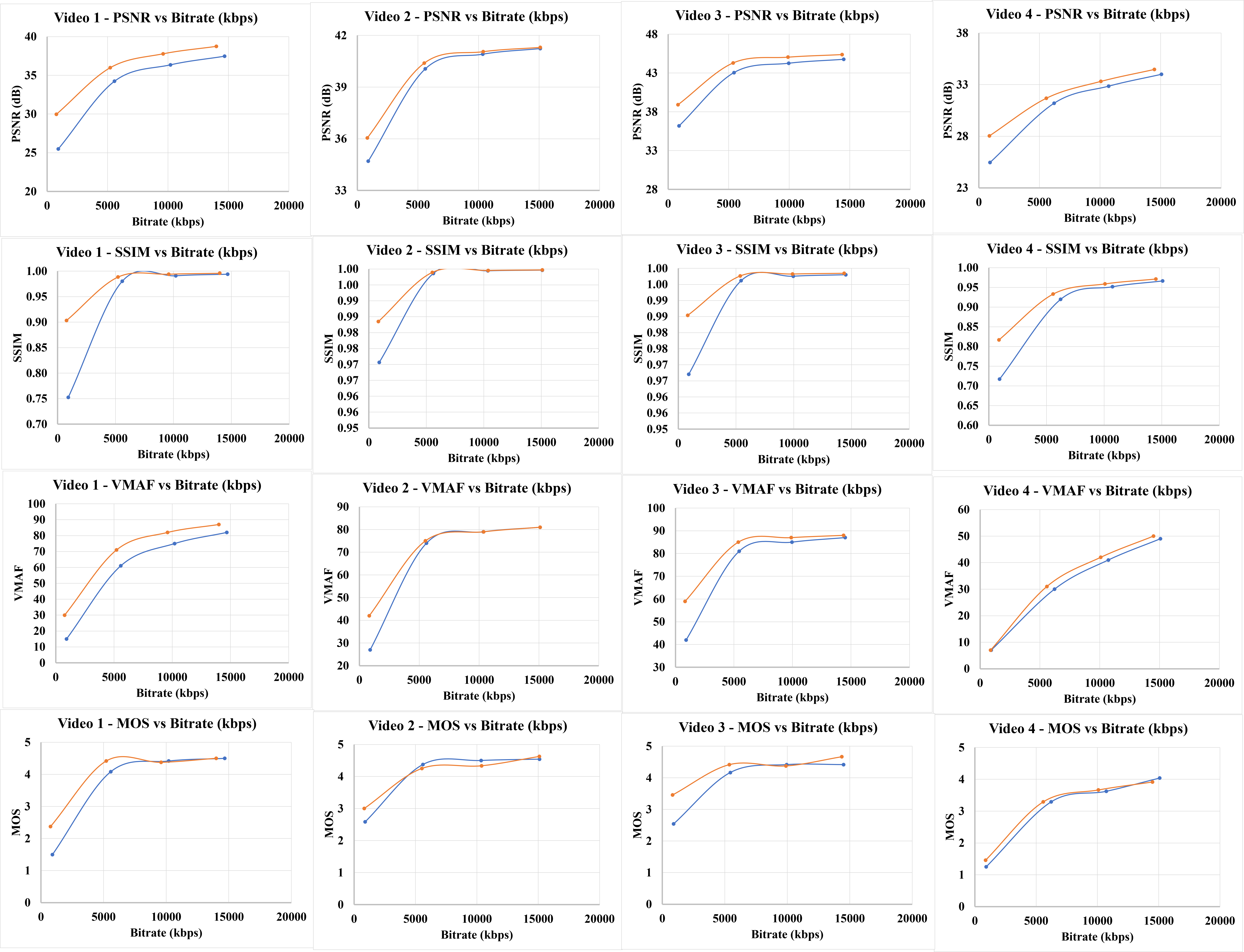}
\caption{RD curves for the four video sequences of the dataset considering four different measurement metrics for distortion:  Row 1, PSNR; Row 2, SSIM; Row 3, VMAF and Row 4, MOS.}
\label{fig:dataset01}
\end{center}
\end{figure*}

\section{Experimental Study of Different Implementations and Variants of BD Metric} \label{sec:ExpStudy}

\subsection{Dataset}

To test the robustness of the different BD functions and implementations, we consider an open source dataset AVT-VQDB-UHD-1~\cite{rao2019Db} consisting of objective and subjective quality metric scores for videos encoded using two state-of-the-art codecs. For this work, we will refer to them as codec A and codec B as our focus here is not on comparing codecs for their compression efficiency gain but rather on studying the methods to do so. The dataset was used in the design and development of ITU-T Rec. P.1204 and is available in~\cite{rao2019Db}. 

For brevity, we consider a subset (four videos) of the full AVT-VQDB-UHD-1 dataset consisting of scores for three objective quality metrics (PSNR, SSIM, and VMAF) and subjective ratings (MOS). Figure~\ref{fig:screenshots} presents the screenshots of the four videos from the AVT-VQDB-UHD-1 dataset. The videos are of 1920x1080p resolution at 60fps, encoded in four different bitrate values, thus resulting in four different test cases. The MOS scores are obtained from subjective tests conducted on a 65$''$ 4K Panasonic screen following ITU-R BT.500-13 recommendations \cite{bt500-14} ensuring reliability and repeatability of the results \cite{rao2019Db}. 

Figure~\ref{fig:dataset01} presents the RD curves for the four videos. It can be observed that the RD curves for the videos are quite different across different quality metrics - from “well-behaved” RD curves to cases where there is a crossover between them. One can also observe that Video 1 and 3 have non-monotonically increasing MOS for codec B, while Video 2 and 4 have monotonically increasing MOS for both codecs. Such test cases represent real-world scenarios and can help us evaluate the performance of various BD functions and implementations for their performance in more practical (industrial and academic) works. This will allow us to identify the challenges and propose recommendations on best practices for using the BD metric, primarily for non-standardization-related activities.

\begin{table*}[t!]
  \centering
  \caption{BD-Rate and BD-Quality results for different implementations considering PSNR, SSIM, and VMAF quality scores for the four cases.}
  \def\arraystretch{1.3}
\begin{tabular}{|lcccccccccccc|}
\hline
\multicolumn{1}{|c|}{\multirow{3}{*}{\textbf{Implementation}}} &
  \multicolumn{6}{c|}{\textbf{Video 01}} &
  \multicolumn{6}{c|}{\textbf{Video 02}} \\ \cline{2-13} 
\multicolumn{1}{|c|}{} &
  \multicolumn{3}{c|}{\textbf{BD-Rate}} &
  \multicolumn{3}{c|}{\textbf{BD-Quality}} &
  \multicolumn{3}{c|}{\textbf{BD-Rate}} &
  \multicolumn{3}{c|}{\textbf{BD-Quality}} \\ \cline{2-13} 
\multicolumn{1}{|c|}{} &
  \multicolumn{1}{c|}{\textbf{PSNR}} &
  \multicolumn{1}{c|}{\textbf{SSIM}} &
  \multicolumn{1}{c|}{\textbf{VMAF}} &
  \multicolumn{1}{c|}{\textbf{PSNR}} &
  \multicolumn{1}{c|}{\textbf{SSIM}} &
  \multicolumn{1}{c|}{\textbf{VMAF}} &
  \multicolumn{1}{c|}{\textbf{PSNR}} &
  \multicolumn{1}{c|}{\textbf{SSIM}} &
  \multicolumn{1}{c|}{\textbf{VMAF}} &
  \multicolumn{1}{c|}{\textbf{PSNR}} &
  \multicolumn{1}{c|}{\textbf{SSIM}} &
  \textbf{VMAF} \\ \hline
\multicolumn{1}{|l|}{\textbf{VCEG-M33}} &
  \multicolumn{1}{c|}{-48.7\%} &
  \multicolumn{1}{c|}{-2.5\%} &
  \multicolumn{1}{c|}{-40.1\%} &
  \multicolumn{1}{c|}{2.62} &
  \multicolumn{1}{c|}{0.04} &
  \multicolumn{1}{c|}{13.01} &
  \multicolumn{1}{c|}{-22.2\%} &
  \multicolumn{1}{c|}{-100.0\%} &
  \multicolumn{1}{c|}{-69.5\%} &
  \multicolumn{1}{c|}{0.65} &
  \multicolumn{1}{c|}{0.00} &
  4.88 \\ \hline
\multicolumn{1}{|l|}{\textbf{JVET-O0003}} &
  \multicolumn{1}{c|}{-50.7\%} &
  \multicolumn{1}{c|}{-56.0\%} &
  \multicolumn{1}{c|}{-45.3\%} &
  \multicolumn{1}{c|}{2.72} &
  \multicolumn{1}{c|}{0.05} &
  \multicolumn{1}{c|}{13.20} &
  \multicolumn{1}{c|}{-28.0\%} &
  \multicolumn{1}{c|}{-34.9\%} &
  \multicolumn{1}{c|}{-25.4\%} &
  \multicolumn{1}{c|}{0.65} &
  \multicolumn{1}{c|}{0.00} &
  4.92 \\ \hline
\multicolumn{1}{|l|}{\textbf{JVET-H0030}} &
  \multicolumn{1}{c|}{-50.7\%} &
  \multicolumn{1}{c|}{-56.0\%} &
  \multicolumn{1}{c|}{-45.3\%} &
  \multicolumn{1}{c|}{2.72} &
  \multicolumn{1}{c|}{0.05} &
  \multicolumn{1}{c|}{13.20} &
  \multicolumn{1}{c|}{-28.0\%} &
  \multicolumn{1}{c|}{-34.9\%} &
  \multicolumn{1}{c|}{-25.4\%} &
  \multicolumn{1}{c|}{0.65} &
  \multicolumn{1}{c|}{0.00} &
  4.92 \\ \hline
\multicolumn{1}{|l|}{\textbf{OS-Python (piecewise)}} &
  \multicolumn{1}{c|}{-50.7\%} &
  \multicolumn{1}{c|}{NaN} &
  \multicolumn{1}{c|}{-45.3\%} &
  \multicolumn{1}{c|}{2.72} &
  \multicolumn{1}{c|}{0.05} &
  \multicolumn{1}{c|}{13.20} &
  \multicolumn{1}{c|}{-28.4\%} &
  \multicolumn{1}{c|}{NaN} &
  \multicolumn{1}{c|}{-25.4\%} &
  \multicolumn{1}{c|}{0.66} &
  \multicolumn{1}{c|}{0.00} &
  4.92 \\ \hline
\multicolumn{1}{|l|}{\textbf{OS-Python (default)}} &
  \multicolumn{1}{c|}{-48.6\%} &
  \multicolumn{1}{c|}{NaN} &
  \multicolumn{1}{c|}{-40.1\%} &
  \multicolumn{1}{c|}{2.61} &
  \multicolumn{1}{c|}{0.05} &
  \multicolumn{1}{c|}{13.01} &
  \multicolumn{1}{c|}{-27.9\%} &
  \multicolumn{1}{c|}{NaN} &
  \multicolumn{1}{c|}{-69.5\%} &
  \multicolumn{1}{c|}{0.67} &
  \multicolumn{1}{c|}{0.00} &
  4.88 \\ \hline
\multicolumn{1}{|l|}{\textbf{OS-Matlab}} &
  \multicolumn{1}{c|}{-48.6\%} &
  \multicolumn{1}{c|}{Inf} &
  \multicolumn{1}{c|}{-40.1\%} &
  \multicolumn{1}{c|}{2.61} &
  \multicolumn{1}{c|}{0.05} &
  \multicolumn{1}{c|}{13.01} &
  \multicolumn{1}{c|}{-27.9\%} &
  \multicolumn{1}{c|}{Inf} &
  \multicolumn{1}{c|}{-69.5\%} &
  \multicolumn{1}{c|}{0.67} &
  \multicolumn{1}{c|}{0.00} &
  4.88 \\ \hline
\multicolumn{1}{|l|}{\textbf{OS-Excel}} &
  \multicolumn{1}{c|}{-48.7\%} &
  \multicolumn{1}{c|}{-2.5\%} &
  \multicolumn{1}{c|}{-40.1\%} &
  \multicolumn{1}{c|}{2.62} &
  \multicolumn{1}{c|}{0.04} &
  \multicolumn{1}{c|}{13.01} &
  \multicolumn{1}{c|}{-22.2\%} &
  \multicolumn{1}{c|}{71.9\%} &
  \multicolumn{1}{c|}{-69.5\%} &
  \multicolumn{1}{c|}{0.65} &
  \multicolumn{1}{c|}{0.00} &
  4.88 \\ \hline
\multicolumn{13}{|c|}{\textbf{}} \\ \hline
\multicolumn{1}{|c|}{\multirow{3}{*}{\textbf{Implementation}}} &
  \multicolumn{6}{c|}{\textbf{Video 03}} &
  \multicolumn{6}{c|}{\textbf{Video 04}} \\ \cline{2-13} 
\multicolumn{1}{|c|}{} &
  \multicolumn{3}{c|}{\textbf{BD-Rate}} &
  \multicolumn{3}{c|}{\textbf{BD-Quality}} &
  \multicolumn{3}{c|}{\textbf{BD-Rate}} &
  \multicolumn{3}{c|}{\textbf{BD-Quality}} \\ \cline{2-13} 
\multicolumn{1}{|c|}{} &
  \multicolumn{1}{c|}{\textbf{PSNR}} &
  \multicolumn{1}{c|}{\textbf{SSIM}} &
  \multicolumn{1}{c|}{\textbf{VMAF}} &
  \multicolumn{1}{c|}{\textbf{PSNR}} &
  \multicolumn{1}{c|}{\textbf{SSIM}} &
  \multicolumn{1}{c|}{\textbf{VMAF}} &
  \multicolumn{1}{c|}{\textbf{PSNR}} &
  \multicolumn{1}{c|}{\textbf{SSIM}} &
  \multicolumn{1}{c|}{\textbf{VMAF}} &
  \multicolumn{1}{c|}{\textbf{PSNR}} &
  \multicolumn{1}{c|}{\textbf{SSIM}} &
  \textbf{VMAF} \\ \hline
\multicolumn{1}{|l|}{\textbf{VCEG-M33}} &
  \multicolumn{1}{c|}{-44.1\%} &
  \multicolumn{1}{c|}{-75.8\%} &
  \multicolumn{1}{c|}{-38.2\%} &
  \multicolumn{1}{c|}{1.76} &
  \multicolumn{1}{c|}{0.01} &
  \multicolumn{1}{c|}{7.50} &
  \multicolumn{1}{c|}{-32.2\%} &
  \multicolumn{1}{c|}{-41.4\%} &
  \multicolumn{1}{c|}{-13.1\%} &
  \multicolumn{1}{c|}{1.20} &
  \multicolumn{1}{c|}{0.04} &
  2.51 \\ \hline
\multicolumn{1}{|l|}{\textbf{JVET-O0003}} &
  \multicolumn{1}{c|}{-50.8\%} &
  \multicolumn{1}{c|}{-53.9\%} &
  \multicolumn{1}{c|}{-47.0\%} &
  \multicolumn{1}{c|}{1.72} &
  \multicolumn{1}{c|}{0.01} &
  \multicolumn{1}{c|}{7.66} &
  \multicolumn{1}{c|}{-33.6\%} &
  \multicolumn{1}{c|}{-39.3\%} &
  \multicolumn{1}{c|}{-12.4\%} &
  \multicolumn{1}{c|}{1.28} &
  \multicolumn{1}{c|}{0.04} &
  2.22 \\ \hline
\multicolumn{1}{|l|}{\textbf{JVET-H0030}} &
  \multicolumn{1}{c|}{-50.8\%} &
  \multicolumn{1}{c|}{-53.9\%} &
  \multicolumn{1}{c|}{-47.0\%} &
  \multicolumn{1}{c|}{1.72} &
  \multicolumn{1}{c|}{0.01} &
  \multicolumn{1}{c|}{7.66} &
  \multicolumn{1}{c|}{-33.6\%} &
  \multicolumn{1}{c|}{-39.3\%} &
  \multicolumn{1}{c|}{-12.4\%} &
  \multicolumn{1}{c|}{1.28} &
  \multicolumn{1}{c|}{0.04} &
  2.22 \\ \hline
\multicolumn{1}{|l|}{\textbf{OS-Python (piecewise)}} &
  \multicolumn{1}{c|}{-50.7\%} &
  \multicolumn{1}{c|}{NaN} &
  \multicolumn{1}{c|}{-47.0\%} &
  \multicolumn{1}{c|}{1.72} &
  \multicolumn{1}{c|}{0.00} &
  \multicolumn{1}{c|}{7.66} &
  \multicolumn{1}{c|}{-33.6\%} &
  \multicolumn{1}{c|}{-34.6\%} &
  \multicolumn{1}{c|}{-12.4\%} &
  \multicolumn{1}{c|}{1.27} &
  \multicolumn{1}{c|}{0.04} &
  2.22 \\ \hline
\multicolumn{1}{|l|}{\textbf{OS-Python (default)}} &
  \multicolumn{1}{c|}{-41.7\%} &
  \multicolumn{1}{c|}{NaN} &
  \multicolumn{1}{c|}{-38.2\%} &
  \multicolumn{1}{c|}{1.75} &
  \multicolumn{1}{c|}{0.00} &
  \multicolumn{1}{c|}{7.50} &
  \multicolumn{1}{c|}{-32.1\%} &
  \multicolumn{1}{c|}{4.6\%} &
  \multicolumn{1}{c|}{-13.1\%} &
  \multicolumn{1}{c|}{1.20} &
  \multicolumn{1}{c|}{0.03} &
  2.51 \\ \hline
\multicolumn{1}{|l|}{\textbf{OS-Matlab}} &
  \multicolumn{1}{c|}{-41.7\%} &
  \multicolumn{1}{c|}{Inf} &
  \multicolumn{1}{c|}{-38.2\%} &
  \multicolumn{1}{c|}{1.75} &
  \multicolumn{1}{c|}{0.00} &
  \multicolumn{1}{c|}{7.50} &
  \multicolumn{1}{c|}{-32.1\%} &
  \multicolumn{1}{c|}{4.6\%} &
  \multicolumn{1}{c|}{-13.1\%} &
  \multicolumn{1}{c|}{1.20} &
  \multicolumn{1}{c|}{0.03} &
  2.51 \\ \hline
\multicolumn{1}{|l|}{\textbf{OS-Excel}} &
  \multicolumn{1}{c|}{-44.1\%} &
  \multicolumn{1}{c|}{-67.4\%} &
  \multicolumn{1}{c|}{-38.2\%} &
  \multicolumn{1}{c|}{1.76} &
  \multicolumn{1}{c|}{0.01} &
  \multicolumn{1}{c|}{7.50} &
  \multicolumn{1}{c|}{-32.2\%} &
  \multicolumn{1}{c|}{-41.4\%} &
  \multicolumn{1}{c|}{-13.1\%} &
  \multicolumn{1}{c|}{1.20} &
  \multicolumn{1}{c|}{0.04} &
  2.51 \\ \hline
\end{tabular}
    \label{tab:allresults}%
\end{table*}%

\vspace{-0.5cm}
\subsection{BD Implementations}
For a comparative evaluation of the different implementations, we consider both the Excel implementations used or proposed in standardization activities (JVT, JCT-VC, and JVET) and commonly used open-source implementations. 

\subsubsection{Standard Implementations}

The following three implementations made available during various standardization activities are used: 
\begin{enumerate}[(a)]
    \item \textit{VCEG-M33} refers to the implementation made available as part of the first contribution by Gisle in April 2001~\cite{VCEG-M33}.
    \item \textit{JVET-O0003} refers to the piecewise cubic interpolation implementation currently used by JVET in standardization activities~\cite{JVET-O0003}.
    \item \textit{JVET-H0030} refers to the extended version of BD function proposed in 2017 by Tourapis et al. used in the default mode (\textit{None}, see Appendix~\ref{h0030Modes} or~\cite{JVET-H0030} for more details on the different modes)~\cite{JVET-H0030}. While the function implemented is the same as that in the JVET-O0003 implementation, the JVET-H0030 implementation supports more than 4 data points and includes additional support for error handling.
\end{enumerate}

\subsubsection{Open Source Implementations}

While there are many different open-source implementations, we consider the below three commonly used implementations, with each representing a different software implementation (MS Excel, MATLAB, and Python):
\begin{enumerate}[(a)]
    \item \textit{OS-Excel }refers to the Bjøntegaard Metric implementation made available in \cite{OS-Excel}. This Excel implementation is quite widely used by many in the industry~\cite{jan_ozer_bd_br,bd_br_ott_verse}.
    \item \textit{OS-Matlab} refers to the MATLAB implementation provided in~\cite{os-matlab} that supports BD calculation with more than 4 data points. 
    \item \textit{OS-Python} refers to the Python implementation available in~\cite{OS-Python}. The implementation has two modes, default and piecewise, represented by OS-Python (default) and OS-Python (piecewise), respectively.
\end{enumerate}

\noindent Notes:
\begin{itemize}
    \item The Matlab code used in~\cite{os-matlab} is an improved MATLAB version of the Bjøntegaard metric~\cite{matlab_gv} with correct integration intervals. The values obtained by the implementation in~\cite{matlab_gv} do not use the recommended integration intervals; hence, the results are not reported here.
    \item There is also another Python implementation, ``BDmetric 0.9.0"~\cite{OS-PythonOther}. However, as the code base is the same, we use only the first implementation to calculate our results, but both implementations should give the same results.
    \item In the OS-Python implementation, the authors “fix” the case when the curve is not monotonic by sorting the metric values. While this provides the additional functionality of not having to sort RD values, it might result in “wrong” results when considering the MOS scores since MOS values 
    at the higher end are not always monotonic (see MOS-Bitrate curve for Video 2 and Video 4 in Fig.~\ref{fig:dataset01}).
\end{itemize}

We present and discuss next the evaluation results of the various implementations. Table~\ref{tab:allresults} presents the results for both BD-Rate and BD-Quality for all four cases for the seven implementations considering three different objective quality metrics - PSNR, SSIM, and VMAF. Unless mentioned otherwise, by BD-Rate and BD-PSNR, we refer to rate and PSNR savings obtained using PSNR as the metric for measuring distortion.

\subsection{BD-Rate and BD-PSNR Results Across Implementations}

Based on the results for different implementations as presented in Table~\ref{tab:allresults}, the following observations can be drawn:
\begin{enumerate}[(a)]
    \item Some open-source implementations \cite{os-matlab, OS-Excel, OS-Python} still use cubic fitting - which might result in “unreliable” results. Since most open-source implementations are provided as it is and practically always reference the original BD contribution, VCEG-M33, it is unclear what kind of fitting (cubic or piecewise-cubic) they support. Hence while using any open-source implementation, it should be made sure that the implementation supports the recommended piecewise-cubic mode.
    \item Even for the same BD function, depending on the nature of RD curves, variation in the results across different implementations can be observed. For example, for all implementations using the cubic fitting, considering Video 2, one can observe that while VCEG-M33 and OS-Excel report a BD-Rate savings of $22.2\%$, OS-Python (default) and OS-Matlab reports a BD-Rate savings of $27.9\%$, which is closer to the value obtained using piecewise-cubic implementations (JVET-O0003, JVET-H0030, and OS-Python (piecewise)). However, the values for implementations using piecewise-cubic fitting (JVET-O0003, JVET-H0030, and OS-Python (piecewise)), the values reported across the four cases are the same (ignoring slight differences due to rounding off errors).
    \item Considering BD-PSNR results, one might observe that the PSNR savings between the two functions (cubic and piecewise-cubic) are more in agreement across respective implementations with each other with some variation as compared to the BD-Rate results, indicating higher stability of BD-PSNR results as compared to BD-Rate results. Thus, reporting BD-PSNR values along with BD-Rate can help better interpret the performance of the compared RD curves.
\end{enumerate}

\textit{Note}: While not explicitly evaluated in this study, it should be noted that VCEG-M33 and JVET-O0003 have support for only 4 data points while other implementations (JVET-H0030, OS-Matlab, OS-Excel, and OS-Python) supports more than 4 data points. The results and hence inferences obtained with RD curves considering more than 4 data points can be somewhat different. 

\begin{table*}[t!]
  \centering
  \caption{Comparison of BD-Rate (MOS) and BD-Quality (MOS) results across different implementations considering subjective (MOS) scores for four different cases.}
\def\arraystretch{1.3}
    \resizebox{1.0\linewidth}{!}{

\begin{tabular}{|l|cccc|cccc|}
\hline
\multicolumn{1}{|c|}{\multirow{2}{*}{\textbf{\begin{tabular}[c]{@{}c@{}}Test \\ Sequence\end{tabular}}}} &
  \multicolumn{4}{c|}{\textbf{BD-Rate (MOS)}} &
  \multicolumn{4}{c|}{\textbf{BD-Quality (MOS)}} \\ \cline{2-9} 
\multicolumn{1}{|c|}{} &
  \multicolumn{1}{c|}{\textbf{\begin{tabular}[c]{@{}c@{}}VCEG-M33\\ (cubic)\end{tabular}}} &
  \multicolumn{1}{c|}{\textbf{\begin{tabular}[c]{@{}c@{}}JVET-O0030 \\ (piecewise-cubic)\end{tabular}}} &
  \multicolumn{1}{c|}{\textbf{\begin{tabular}[c]{@{}c@{}}JVET-H0030 \\ (piecewise-cubic)\end{tabular}}} &
  \textbf{SCENIC} &
  \multicolumn{1}{c|}{\textbf{\begin{tabular}[c]{@{}c@{}}VCEG-M33\\ (cubic)\end{tabular}}} &
  \multicolumn{1}{c|}{\textbf{\begin{tabular}[c]{@{}c@{}}JVET-O0030\\ (piecewise-cubic)\end{tabular}}} &
  \multicolumn{1}{c|}{\textbf{\begin{tabular}[c]{@{}c@{}}JVET-H0030\\ (piecewise-cubic)\end{tabular}}} &
  \textbf{SCENIC} \\ \hline
Video 1 &
  \multicolumn{1}{c|}{NaN} &
  \multicolumn{1}{c|}{-43.8\%} &
  \multicolumn{1}{c|}{-44.4\%} &
  -59.5\% &
  \multicolumn{1}{c|}{NaN} &
  \multicolumn{1}{c|}{0.60} &
  \multicolumn{1}{c|}{0.60} &
  0.80 \\ \hline
Video 2 &
  \multicolumn{1}{c|}{-0.2\%} &
  \multicolumn{1}{c|}{-2.2\%} &
  \multicolumn{1}{c|}{-2.2\%} &
  8.1\% &
  \multicolumn{1}{c|}{0.18} &
  \multicolumn{1}{c|}{0.04} &
  \multicolumn{1}{c|}{0.04} &
  -0.02 \\ \hline
Video 3 &
  \multicolumn{1}{c|}{NaN} &
  \multicolumn{1}{c|}{NaN} &
  \multicolumn{1}{c|}{NaN} &
  -50.4\% &
  \multicolumn{1}{c|}{NaN} &
  \multicolumn{1}{c|}{NaN} &
  \multicolumn{1}{c|}{NaN} &
  0.42 \\ \hline
Video 4 &
  \multicolumn{1}{c|}{-32.2\%} &
  \multicolumn{1}{c|}{-10.9\%} &
  \multicolumn{1}{c|}{-10.9\%} &
  -19.1\% &
  \multicolumn{1}{c|}{0.12} &
  \multicolumn{1}{c|}{0.12} &
  \multicolumn{1}{c|}{0.12} &
  0.20 \\ \hline
\end{tabular}
  }
  \label{tab:mos}%
\end{table*}%

\subsection{BD-Rate and BD-Quality Results Across Quality Metrics} \label{subsec:DiffQualities}

\begin{enumerate}[(a)]
    \item Considering SSIM as the objective quality metric, one can observe quite a lot of variation (disagreement) across different implementations. For example, in Video 4, OS-Python (default) and OS-Matlab indicate that codec B performs better than codec A (positive BD-Rate savings) as compared to other implementations indicating better performance for codec A (over codec B). In other cases, many implementations do not agree on the savings figures. It is also interesting to note that, considering the magnitude of BD-Quality (SSIM) results, the BD-Rate (SSIM) results look non-realistic for most of the implementations.
    \item Considering VMAF, for both BD-Rate and BD-Quality savings, one can observe that the results are the same across different implementations for the same BD function. However, values can vary depending on whether the implementation uses cubic or piecewise cubic fitting.
    \item Comparing the values across the exact implementation, one can observe that the BD-Rate and BD-Quality savings can vary a lot depending on the choice of the quality metric. For example, for Video 4, JVET-O0003 indicates a BD-Rate savings of $-33.6\%$ and $-12.4\%$ considering PSNR and VMAF as quality metrics, respectively. A possible reason behind this could be the range of actual bitrate and quality overlap as discussed earlier in ~\ref{sec:overlapMetric}.
\end{enumerate}

\subsection{BD-Rate and BD-Quality Results Considering Subjective (MOS) Scores}

In addition to the three BD implementations from standardization activities, for this study, we additionally consider the Subjective Comparison of ENcoders based on fItted Curves (SCENIC) metric~\cite{scenic} which computes the average bitrate and MOS difference between two RD curves considering subjective (MOS) scores. The basic argument behind the proposed metric is that, since MOS is not a linear metric, a non-symmetrical function should be used to map bit rate values to MOS. In Table~\ref{tab:mos}, we report BD-Rate (MOS) and BD-Quality (MOS) results for the two BD functions and their corresponding implementations and SCENIC (using the open-source implementation available in~\cite{scenic_github}) for both rate and MOS savings. For SCENIC calculations, we used the MATLAB-based open-source implementation of the metric made available in~\cite{scenic_github}. 

From Table~\ref{tab:mos} it can be seen that for BD-Rate calculation, considering all four cases, none of the three BD functions reaches an agreement with entirely different values and, in some cases, even indicates a contrasting performance of the codecs. For example, for Video 2, while the first three indicate a better performance of codec B, the SCENIC metric reports a better performance of codec A. However, for the SCENIC metric, considering the MOS-Bitrate curves presented in Figure~\ref{fig:dataset01}, the BD-Rate (MOS) values do not seem very realistic. In contrast, the BD-Quality (MOS) values do look more realistic. However, in the absence of ground truth of actual bitrate or quality savings, one cannot truly quantify the correctness of either of the metrics.

\section{Observations and Recommendations} \label{sec:obser}

\subsection{Discussion on Suitability of Metrics Other than PSNR}

Our results using two additional objective quality metrics, SSIM and VMAF, indicate that the BD metric computation for metrics other than PSNR should be done with caution. The values of SSIM, given its highly nonlinear nature with bitrate, vary by very small magnitudes, especially in the mid-high bitrate ranges. Hence, using SSIM as the quality metric for BD calculations results in unreliable results (see Table~\ref{tab:allresults}). Similarly, values obtained using MOS as the quality metric can often be unreliable due to possible saturation at higher bitrates or crossover due to the non-monotonic nature and overlapping confidence intervals. To address this, authors in \cite{ugur2010high} use lower order polynomial for curve-fitting of five operational MOS-Bitrate points to avoid overfitting to minor MOS variations resulting in more accurate BD-Rate calculations. 

Similarly, alternative approaches to using SSIM-based BD measurements can be a change of the scale (e.g., 1-SSIM or 1/SSIM ~\cite{dai2014ssim}) or using SSIM-based distortion metrics as a function of MSE between source and reconstructed signals and source signal variance~\cite{yeo2013rate}. However, such possible alternatives (change of scale, fitting, or calculation of BD values separately for different quality ranges) must be further evaluated and compared for accuracy and suitability for different applications, which we leave for future work.

\subsection{Key Observations}

Based on the results presented in Section~\ref{sec:ExpStudy}, the following key observations can be reported:
\begin{enumerate}[(a)]
    \item Many open-source implementations still use cubic fitting instead of the recommended piecewise-cubic fitting for interpolation between the RD points.
    \item BD-Rate and BD-Quality savings can vary depending on whether the implementation uses cubic-fitting or piecewise cubic.
    \item Depending on the choice of quality metric, the corresponding BD-Rate, and BD-Quality savings figure can vary significantly.
    \item When using quality metrics other than PSNR, especially SSIM, the results need to be interpreted with caution!
    \item BD savings figures using MOS as the metric for distortion can be pretty unstable and hence, unreliable.
    \item BD-Rate and BD-Quality savings figures do not necessarily agree across different metrics.
\end{enumerate}

\subsection{Recommendations}
Based on the key observations discussed above, we present next a list of important recommendations.
\begin{enumerate}[(a)]
    \item BD-Rate values are quite sensitive to the nature of RD curves (overlap). Ideally, computation of BD-Rate should be limited to only cases where the RD curves are “well-behaved.” When that is not the case, the BD-Rate values should be supplemented by RD curves and additional data (e.g., BD-Quality savings) for better comparison and interpretation.
    \item The implementation used should use piecewise-cubic interpolation for curve fitting. Our evaluation observed that JVET-O0003, JVET-H0030, and OS-Python (piecewise) implementations provide the correct function implementation. However, agreement of values obtained using a piecewise-cubic fitting with values obtained using cubic fitting can be used as an additional check on the reliability of the obtained savings figures.
    \item When using quality metrics other than PSNR values (especially SSIM and MOS), the results should be interpreted with caution. 
    \item MOS-based BD values should be calculated when MOS is monotonically increasing, and the RD curves should be “well-behaved” (non-overlapping confidence interval, etc.).
    \item Not all implementations support more than 4 data points. Hence, if using more than four data points for BD calculation, ensure that the implementation supports that (JVET-H0030 or the OS-Python (piecewise) implementation).
\end{enumerate}

\subsection{On the Reciprocity of BD-Rate and BD-Quality Savings}

Additionally, one must consider that one of the reasons behind using the log of Bitrate values for BD-Rate calculation is that, otherwise, during the calculation of bitrate savings, a higher bitrate saving is obtained at the high bitrate end. The use of the log bitrate scale results in linear curves for the two quality-bitrate curves. The reciprocity of calculation of BD-Rate and BD-PSNR seems to work because, in this case, both PSNR and Rate are on the log scale. The question remains, however, with the use of quality metrics other than PSNR for BD calculations, does the reciprocity remain valid? For example, when considering SSIM, as discussed in ~\ref{subsec:DiffQualities}, the magnitude of BD-Quality (SSIM) savings figures do not agree with the savings figures reported by BD-Rate (SSIM). 

Another critical factor to consider is that while the initial metric PSNR was unbounded, the other metrics such as SSIM, VMAF, and MOS are not. As observed in our studies and also argued by the authors in~\cite{scenic} and ~\cite{ugur2010high}, the saturation of the quality metrics at a higher bitrate range and the shape of the RD curves need to be considered for the calculation of the BD metric values. This is more relevant now, considering that most services already target high-quality ranges. Hence, a method that considers the actual operating range can help one obtain more “practical” savings figures. Also, considering MOS as the quality metric, our results in Table~\ref{tab:mos} indicate that when MOS scores are not monotonically increasing and/or when there is a crossover between the RD curves, the values obtained for BD-Rate (MOS) and BD-Quality (MOS) can be very misleading (especially, if reported without mentioning the actual measurement values and RD curves).

\section{Possible future extensions} \label{sec:possibleExtension}

\subsection{Extensions of BD-BR for Learning-based Metrics}
The field of image and video quality assessment has been rapidly evolving. Over the past 20 years, we have seen a proliferation of newer quality metrics from SSIM, and VMAF discussed earlier to new deep learning-based metrics such as \cite{LPIPS,Bosse-WQDIQAM,DBCNN}. Due to different encoding and streaming requirements, there is also a growing interest and work towards the development of application-specific metrics such as the ones proposed in \cite{barmanMLmodel,demi,nofu,Nasim_GAN,saman_NRGVQM} for spectator and cloud gaming applications. Also, recently there has been a growing interest and work towards assisted, or standalone Artificial Intelligence (AI) based video compression \cite{AIVC,GoogleGAN}. In order to measure coding efficiency gains, operators and service providers of such applications would ideally want to use custom AI-based metrics for estimating the rate and quality savings, as was observed in much of the literature discussed in Table~\ref{tab:CodecComparisonLiterature}. Hence, considering the fact that depending on the nature of such applications (live vs. on-demand, gaming vs. non-gaming), the operational range (QP or bitrate values) could vary significantly \cite{Barman2017QoMEX17,barmanMLmodel}, there is a need for more advanced and generic methods to compute average delta estimates. We discuss next the possible approach based on weighted average or network-density-based average modifications. 

\subsection{Network-aware Extensions of the BD-BR Method}

By looking back at BD-Quality definition in Section~\ref{sec:bd-psnr}, we note, that this quantity can also be understood as: BD-Quality = $$ \bar{Q}_A(U_{R_{min},R_{max}}) - \bar{Q}_B(U_{R_{min},R_{max}})$$ 
where $\bar{Q}_A(p)$ and $\bar{Q}_B(p)$ denote an average value of functions $\bar{Q}_A(R)$ and $\bar{Q}_B(R)$ respectively:
\begin{equation}  \label{eqn:netavg}
 \resizebox{.9\hsize}{!}{$
  \bar{Q}_A(p) = \int {Q}_A(R)p(R)dR, 
    \bar{Q}_B(p) = \int {Q}_B(R)p(R)dR,
    $}
\end{equation}
where $R$ is treated as a random variable with known probability mass function $p$: $R \sim p$.

In its original definition, the BD method employs simple uniform distribution as the basis for averaging:
\begin{equation} \label{eqn:netbdbr}
\resizebox{1.0\hsize}{!}{$
  p(R) = U_{R_{min},R_{max}}(R) = 
  \begin{cases}
    0       & \quad R < R_{min} \\
    \frac{1}{R_{max}-R_{min}} & \quad R_{min} \leq R \leq R_{max}. \\
    0 & \quad R > R_{max}
  \end{cases}$}
\end{equation}

In cases when network distributions $p(R)$ are known, it could be further argued that network-pdf-weighted average quantities (Eqn~\ref{eqn:netavg}) would provide a much more relevant assessment of codec performance compared to the uniform-density average. And hence, for such applications, it may make more sense to use network-weighted BD-Quality estimates:
\begin{equation}
    BD-Quality(p) = \bar{Q}_A(p) - \bar{Q}_B(p)
\end{equation}
where $p(R)$ defines a network pdf model to be used for analysis, and where $Q_A(R)$ and $Q_B(R)$ are model quality-rate functions produced by interpolating sample points, the same way as in the original BR method. In other words, as the range of applications of the BD method broadens, we may anticipate this method to be extended and used not only with different quality metrics but also with densities used to average the results.

\begin{figure}[t!]
\begin{center}
\includegraphics[width=1.0\linewidth]{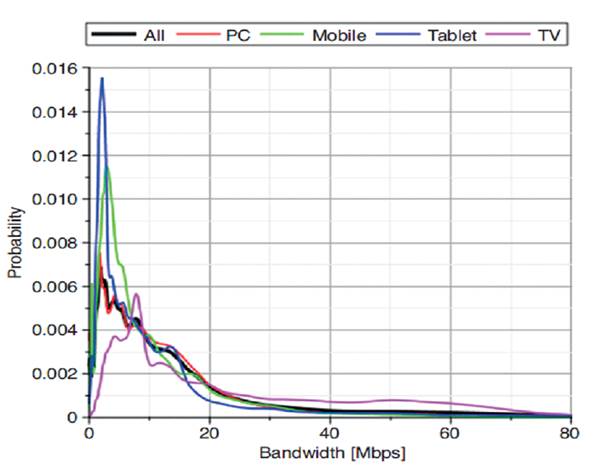}
\caption{Example of network distributions measured for different receiving devices of streaming services. Reproduced from~\cite{Reznik_OptimalDesign}.}
\label{fig:network}
\end{center}
\end{figure}


\section{Conclusions} \label{sec:conclusion}

In this work, we first presented a tutorial about the principles of codec performance comparison and the BD method for computing the average codec performance gains between different codecs. This was followed by a detailed discussion of the history and evolution of the BD metric, its newer variants, and different open-source implementations. We performed an experimental study to evaluate the various open-source implementations of the BD method and its variants. It was found that, depending on the implementation used, for the same dataset, different values can be obtained. This is primarily due to the use of the “deprecated” cubic fitting instead of the recommended piecewise-cubic fitting function. When using quality metrics other than PSNR, our results also showed that metrics such as SSIM and MOS might result in very unstable (and often unrealistic) results. Considering MOS, due to its possible non-monotonic nature, all metrics provide different results, and hence it is challenging to agree on a particular “savings” figure, be it bitrate or MOS. Based on the results, critical observations and a set of recommendations were provided. In short, unless the RD curves for the two codecs compared are “well behaved,” BD results should be interpreted with caution and supported with additional measurements such as BD-Quality savings and RD plots. BD values for metrics other than PSNR should be reported and interpreted cautiously. 

While the BD metric is quite simple and provides a good indication of relative savings, many limitations exist. One such limitation was discussed using an example case study with a crossover between two RD curves. It was shown that depending on the selected bitrate range; the BD metric fails to capture the actual codec performance. Also, given the era of newer quality metrics and other content such as HDR and Point Cloud, many opportunities exist to design more advanced metrics, either as an enhancement of existing BD metrics or alternative approaches. One such alternative approach for improved BD metric design taking into account a more realistic operations range by considering network-density-based average modifications was discussed. 

\section*{Acknowledgements}

Nabajeet Barman would like to thank Alexis M. Tourapis (Apple Inc.) for his initial guidance and clarifications on the JVET-H0030 contribution. Nabajeet Barman would also like to thank the authors (Rakesh Rao and Steve G\"{o}ring from TU Ilmenau, Germany) of the AVT-VQDB-UHD-1 dataset used in this work for their help in resolving queries regarding the dataset.

\printbibliography

\begin{IEEEbiography}[{\includegraphics[width=1in,height=1.25in,clip,keepaspectratio]{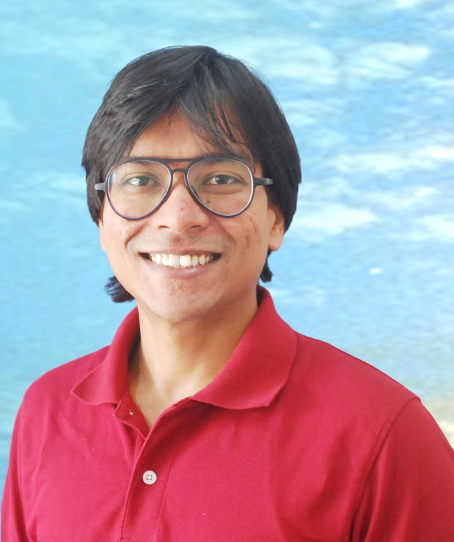}}]{Nabajeet Barman} (M'19) is currently a Senior Research Scientist at Sony Interactive Entertainment (PlayStation) working  on intelligent content and context aware encoding strategies and perceptual quality assessment for their cloud gaming application. Previously he was a Principal Video Systems Engineer, Research at Brightcove where he worked on optimal video encoding strategies, perceptual quality assessment, and end-to-end optimization. He is also a Fellow of the Higher Education Academy (FHEA), UK. He received his PhD and MBA from Kingston University, London, an MSc in IT from Universität Stuttgart, Germany, and B.Tech in Electronics Engineering from NIT, Surat, India. Previously, he was a Lecturer in Applied Computer Science (Data Science) at Kingston University, London. From 2012-2015, he worked in different capacities across various industries including Bell Labs, Stuttgart, Germany after which he joined Kingston University as a Marie Curie Fellow with MSCA ITN QoE-Net from 2015 to 2018, and a Post-Doctoral Research Fellow from 2019-2020. He is a Board Member of the Video Quality Expert Group (VQEG) where he chairs the Computer Generated Imagery and Emerging Technology Group. He has published in many international conferences and journals and is an active reviewer for many conferences and journals.
\end{IEEEbiography}
\begin{IEEEbiography}[{\includegraphics[width=1in,height=1.25in,clip,keepaspectratio]{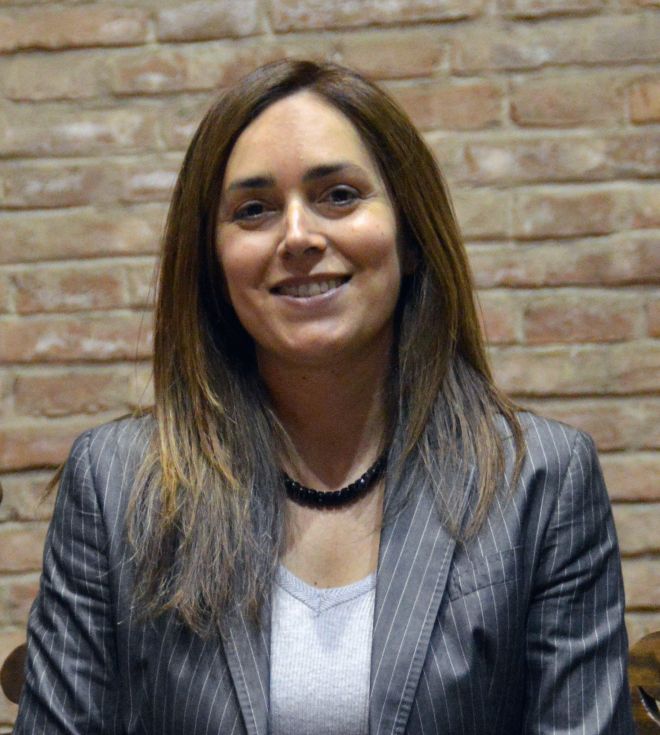}}]{Maria Martini}
 [SrM’07] is Professor in the Faculty of Science, Engineering and Computing at Kingston University, London, U.K., where she also leads the Wireless Multimedia Networking Research Group and she is the Course Director for the MSc in “Networking and Data Communications.” She is a Fellow of The Higher Education Academy (HEA). She received the Laurea degree in electronic engineering (summa cum laude) from the University of Perugia, Italy, in 1998 and the PhD degree in Electronics and Computer Science from the University of Bologna, Italy, in 2002. She has led the KU team in a number of national and international research projects, funded by the European Commission (e.g., OPTIMIX, CONCERTO, QoE-NET, Qualinet), U.K. research councils (e.g., EPSRC, British Council, Royal Society), Innovate UK, and international industries.  Associate Editor for IEEE Signal Processing Magazine (2018-2021) and  IEEE Transactions on Multimedia (2014-2018), she was lead guest editor for the IEEE JSAC special issue on "QoE-aware wireless multimedia systems" (2012), and  editor for IEEE Journal of Biomedical and Health Informatics (2014), IEEE Multimedia (2018),  Int. Journal on Telemedicine and Applications, among others. Expert Evaluator and Panel Member for the European Commission and for national funding agencies (e.g. EPSRC in the UK), she  is part of the NetWorld2020 ETP Expert Group, Board member of the Video Quality Expert Group (VQEG) and member of the IEEE Multimedia Communications technical committee, currently serving in the Awards Committee and having served as vice-chair (2014-2016), as chair  of the 3D Rendering, Processing, and Communications Interest Group (2012-2014), key member of the QoE and multimedia streaming IG. Her research interests include wireless multimedia networks, video quality assessment, decision theory, machine learning, and medical applications. She authored about 200 international scientific articles and  book sections,  international patents and contributions to international standards (IEEE and ITU). She currently chairs the IEEE P3333.1.4 standardization working group on the quality assessment of light field imaging. 

\end{IEEEbiography}

\begin{IEEEbiography}[{\includegraphics[width=1in,height=1.25in,clip,keepaspectratio]{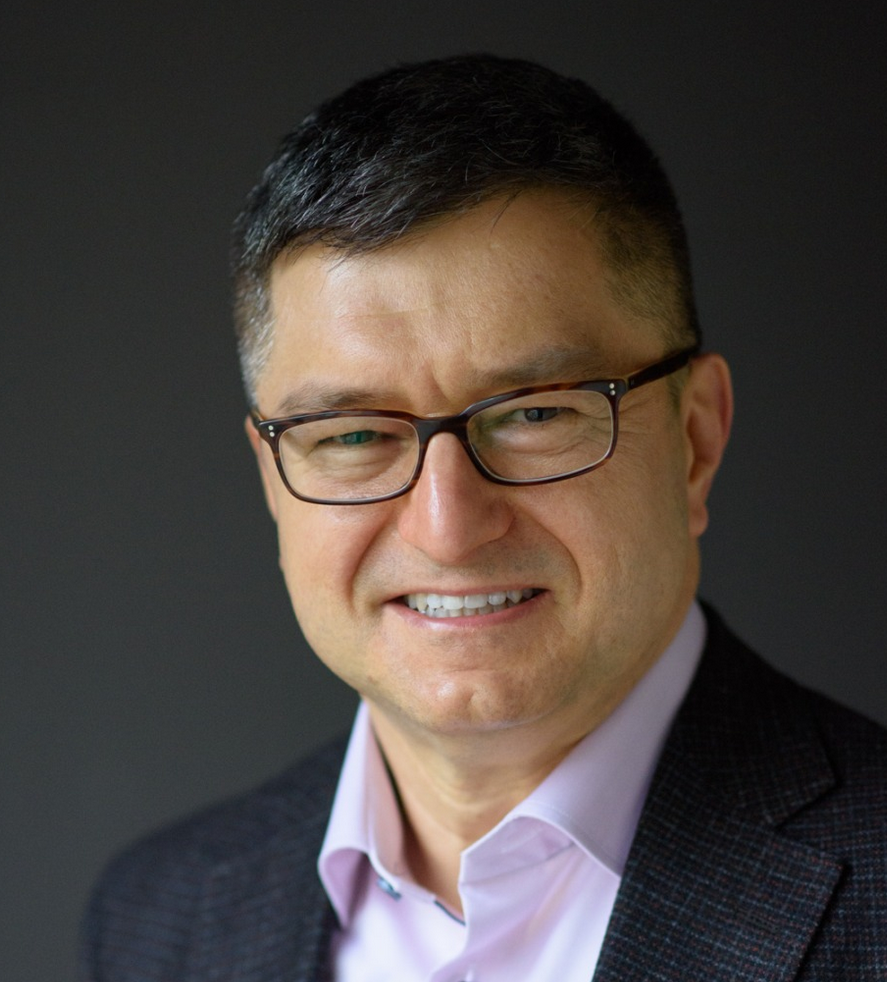}}]{Yuriy Reznik} is a technology fellow and the vice president of research at Brightcove, Inc., Boston, MA. Previously, he held engineering and management positions at InterDigital, San Diego, CA, from 2011 to 2016, Qualcomm, San Diego, from 2005 to 2011, and RealNetworks, Seattle, WA, from 1998 to 2005. In 2008, he was a visiting scholar at Stanford University, Stanford, CA. Since 2001, he has also been involved in the work of the ITU-T SG16 and MPEG standards committees and has made contributions to several multimedia coding and delivery standards, including ITU-T H.264/MPEG-4 AVC, MPEG-4 ALS, ITU-T G.718, ITU-T H.265/MPEG HEVC, and MPEG-DASH. Several technologies, standards, and products that he has helped to develop (RealAudio/RealVideo, ITU-T H.264/MPEG-4 AVC, Zencoder, and Brightcove CAE) have been recognized by the NATAS Technology \& Engineering Emmy Awards. He holds a PhD degree in computer science from Kyiv University, Kyiv, Ukraine. He is a senior member of IEEE and SPIE and a member of the ACM, AES, and SMPTE. He is a co-author of over 120 conference and journal papers and co-inventor of over 80 granted US patents.
\end{IEEEbiography}

\vskip -2\baselineskip plus -1fil

\appendices

\section{Summary of Different modes supported in JVET-H0030} \label{h0030Modes}
The JVET-H0030 contribution supports seven different modes, which are set using the optional parameter \textit{bMode} for the BD-Rate case. Extrapolation, when used, is performed via a linear extrapolation in the log bitrate domain (while interpolation between the data points is piecewise cubic). The different modes supported are:
\begin{itemize}
    \item “None”: The default mode wherein no extrapolation is considered. However, if no overlap exists, the function reports either $-100\%$ or $100\%$ BD-Rate change depending on the location relationship of the two curves.
\item “Low”: Adaptive extrapolation is performed only and only if there is no overlap and only for the “higher” performance curve towards the low PSNR end.
\item “High”: Adaptive extrapolation is performed only and only if there is no overlap and only for the “lower” performance curve towards the higher PSNR end.
\item “Both”: Adaptive extrapolation is performed only and only if there is no overlap, and for both curves achieving maximal coverage.
\item “LowAlways”: Extrapolation is always performed for the “higher” performance curve towards the low PSNR end. 
\item “HighAlways”: Extrapolation is always performed for the “lower” performance curve towards the higher PSNR end. 
\item “BothAlways”: Extrapolation is performed for both curves achieving maximal coverage.
\end{itemize}

\end{document}